 \documentclass[acmtog]{acmart}
  \acmSubmissionID{papers\_566}
  
\AtBeginDocument{%
  \providecommand\BibTeX{{%
    \normalfont B\kern-0.5em{\scshape i\kern-0.25em b}\kern-0.8em\TeX}}}

\usepackage{array,multirow}
\usepackage{tabularx}
\usepackage{enumitem}

\begin{document}

\title[Moving Avatars and Agents in XR]{Moving Avatars and Agents in Social Extended Reality Environments}

\author{Jann Philipp Freiwald}
\email{jann.philipp.freiwald@uni-hamburg.de}
\affiliation{%
  \institution{Universit\"at Hamburg}
  \country{Germany}
}

\author{Susanne Schmidt}
\email{susanne.schmidt@uni-hamburg.de}
\affiliation{%
    \institution{Universit\"at Hamburg}
    \country{Germany}
}

\author{Bernhard E. Riecke}
\email{ber1@sfu.ca}
\affiliation{%
    \institution{Simon Fraser University}
    \country{Canada}
}

\author{Frank Steinicke}
\email{frank.steinicke@uni-hamburg.de}
\affiliation{%
    \institution{Universit\"at Hamburg}
    \country{Germany}
}

\renewcommand{\shortauthors}{Freiwald et al.}

\begin{abstract}

Natural interaction between multiple users within a shared virtual environment (VE) relies on each other's awareness of the current position of the interaction partners.
This, however, cannot be warranted when users employ noncontinuous locomotion techniques, such as teleportation, which may cause confusion among bystanders.

In this paper, we pursue two approaches to create a pleasant experience for both the moving user and the bystanders observing that movement.
First, we will introduce a \emph{Smart Avatar} system that delivers continuous full-body human representations for noncontinuous locomotion in shared virtual reality (VR) spaces.
\emph{Smart Avatars} imitate their assigned user's real-world movements when close-by and autonomously navigate to their user when the distance between them exceeds a certain threshold, i.e., after the user teleports.
As part of the \emph{Smart Avatar} system, we implemented four avatar transition techniques and compared them to conventional avatar locomotion in a user study, revealing significant positive effects on the observers' spatial awareness, as well as pragmatic and hedonic quality scores.

Second, we introduce the concept of \emph{Stuttered Locomotion}, which can be applied to any continuous locomotion method.
By converting a continuous movement into short-interval teleport steps, we provide the merits of noncontinuous locomotion for the moving user while observers can easily keep track of their path.
Thus, while the experience for observers is similarly positive as with continuous motion, a user study confirmed that \emph{Stuttered Locomotion} can significantly reduce the occurrence of cybersickness symptoms for the moving user, making it an attractive choice for shared VEs.
We will discuss the potential of \emph{Smart Avatars} and \emph{Stuttered Locomotion} for shared VR experiences, both when applied individually and in combination.
\end{abstract}



\maketitle

\section{Introduction}
\label{sec:Introduction}
Social virtual reality (VR) platforms such as the metaverse received enormous attention recently.
In these virtual environments (VEs), multiple users can meet and interact with each other via their virtual representations, so-called avatars.
In order for users and their avatars to travel, state-of-the-art consumer VR headsets typically employ controller-based and/or teleportation-based locomotion techniques.
Controller-based input, for example via a joystick, produces continuous camera movement that does not correlate with the user's physical movement.
The resulting sensory conflict between the visual and vestibular senses constitutes the most widely accepted theory for the cause of cybersickness~\cite{laviola2000discussion}.
On the other hand, noncontinuous teleportation has been shown to cause significantly less cybersickness~\cite{Buttussi2019LocomotionInPlace}, thus offering an efficient and convenient way to move through VEs, especially over long distances.
Notwithstanding this advantage of noncontinuous over continuous locomotion in terms of the experience of the moving user, there is a yet rarely studied limitation to teleporting in shared environments.
When avatars abruptly teleport between different locations, it is irritating and challenging for observers to keep track of their movements and trajectories.
In this context, \citet{freiwald2021co-presence} showed that noncontinuous visualizations lead to a reduction in co-presence and perceived fairness in a competitive game context.
This results in a discrepancy between the needs of observers, who would benefit from the high predictability of continuous movements (e.g., joystick, motion-based control), and the moving user themselves, who may have a more pleasant VR experience when performing noncontinuous movements (e.g., teleportation, jumping, dashing).\\

In this paper, we approach this mismatch of user requirements from two directions.
First, we address the problem from an observer perspective, by developing four novel techniques to visualize noncontinuous movements (e.g., resulting from teleportation) as continuous transitions through an advanced avatar system.
Second, we consider the problem from a first-person perspective, by converting conventional continuous movements (e.g., resulting from joystick input) into a series of short-distance teleport steps to reduce cybersickness.
All introduced techniques are compared in a two-stage user study and are bundled in a toolkit that is publicly available via~\citet{SmartAvatarsGithub}.\\

\section{Related Work}
\label{sec:RelatedWork}
Our work builds upon two areas of VR research: (i) avatar visualization and (ii) locomotion.
Here, we will focus on findings related to measures such as spatial awareness, embodiment, presence and cybersickness.

\subsection{Avatar Visualizations}
The visual representation of any locomotion technique is usually bound to the avatar of the moving user.
In this context, \citet{steed2016wild} demonstrated that the sense of presence and embodiment can be improved by providing users with self-avatars that represent their real body.
\citet{Waltemate2018PersonalizedAvatars} further reported that personalized humanoid avatars made through photogrammetry received favorable ratings regarding body ownership and sense of presence compared to generic humanoid avatars.
Previous work has further shown that displaying a virtual body improves the user's overall distance estimation, as it gives a sense of scale and relation~\cite{Mohler2010AvatarDistanceEstimation}.
Regarding discrepancies between the positioning of the user and their avatar, an experiment conducted by \citet{perez2012bodyconnected} showed that the alignment between the user's real arms and their virtual representations, as well as the distance between them, were less important for eliciting a body ownership illusion than other factors, such as synchronous visual-tactile stimulation.\\

While an avatar beyond visible hands is preferable but mostly optional for single-user VR experiences, it is mandatory to enable effective spatial relations and interactions between multiple users in shared VR spaces.
In two multi-user interaction studies, Pan and Steed found that embodiment via self-avatars had a significant positive effect on trust formation as well as task performance in cooperative collaboration tasks \cite{pan2017impact,pan2019foot}.
Furthermore, \citet{casaneuva2001presence} showed evidence that gestures and facial expressions, or realistic human-like avatars in general, can increase co-presence in multi-user scenarios.
\citet{freiwald2021co-presence} also found that the visual representation of a user's locomotion is a stronger factor for co-presence in shared VR spaces than avatar appearance.
As part of a motion-based input technique, \citet{freiwald2020VRStrider} introduced a full-body avatar system that maps cycling biomechanics of the user's legs to a human gait animation.
Other options for full-body avatar systems include approximate realistic animations using inverse kinematics~\cite{grochow2004style} and full-body tracking~\cite{sra2015metaspace}.
Furthermore, several VR online platforms offer full-body avatar animation systems that are compatible with both inverse kinematics and full-body tracking solutions.
For example, VRChat's current Avatar 3.0 implementation uses Unity's Mecanim animation system to provide its users with a customizable animation state machine with locomotion, gesture, and inverse kinematic layers~\cite{VRChat2022Avatars30}.

Lastly, avatar systems are not limited to shared VR spaces.
Besides typical non-immersive online games and social experiences, they can also be employed in mixed reality remote setups~\cite{Piumsomboon2018mini-me}.
Here, telepresence systems such as Holoportation or Remote Fusion allow the use of real-time 3D human reconstructions in addition to generic or procedural avatar animations~\cite{orts2016holoportation,adcock2013remotefusion,teo2019mixed}.
Their drawback, however, is often a lack of plausible locomotion visualizations for those human representations.\\

The previously presented related work mainly addresses the visual representation of avatars in single- or multi-user VEs, without explicitly studying correlations with locomotion.
In particular, the challenges posed by noncontinuous user movements in shared VEs have not yet been studied in depth.
Spatial awareness of other users in shared virtual spaces is a necessity which is reflected in the usability ISO norm 9241-11. Unexpected behavior should be avoided as it leads to aversion, which also applies to the avatars of others. For example, when an avatar unexpectedly teleports directly into one’s personal space, there is no time to avoid this objectionable situation.
We therefore introduce a system to create awareness of a user’s imminent movement and to potentially delay their visualizations to enable observers to react preemptively.

\subsection{Locomotion}

In VR, the choice of locomotion has great impact on the user experience and interaction paradigms in general.
Different techniques influence the interaction dimensions of immersion, flow, manageability, sense of effectiveness, and psychophysical discomfort (cybersickness) to varying degrees~\cite{boletsis2019vr}.
In a systematic literature review, \citet{boletsis2017new} categorized locomotion into four distinct types: motion-based, room-scale-based, controller-based and teleportation-based.

Concrete implementations of the motion-based type are exemplified in works such as the \emph{Myo Arm-Swinging}~\cite{mccullough2015myo}, \emph{Accelerometer Walking-in-Place}~\cite{wilson2016vr}, \emph{VR Strider} \cite{freiwald2020VRStrider}, \emph{Wii-Leaning}~\cite{harris2014human}, \emph{LMTravel}~\cite{cardoso2016comparison}, and \emph{VR-STEP}~\cite{tregillus2016vr}.
\citet{ferracani2016locomotion} introduced locomotion via a metaphorical hand gesture that mimics the control of a lever by pushing the hand forward.
However, this technique does not support continuous motion along a curve, as the pushing gesture initiates a forward movement that has to be explicitly stopped by performing another gesture.

Controller- and teleportation-based locomotion represent the current default options for consumer VR headsets like the HTC Vive or Oculus Quest.
A key difference between the two approaches is the avatar's movement between the starting point and the destination, which is either continuous or noncontinuous in conventional controller- or teleportation-based locomotion, respectively.
A few projects have explored how to combine the two approaches by introducing discrete steps for controller-based locomotion on the one hand, and by creating a continuous avatar representation for teleportation on the other hand~\cite{folmer2021teleportation}.

\citet{farmani2020evaluating} presented discrete viewpoint control methods, which either rotate the user's viewpoint by a fixed degree or translate the user's viewpoint by a fixed distance.
The translations are limited to forward/backward movements of each 1 meter, which are triggered through mouse buttons.
Their study results suggest that both discrete rotations and discrete lateral movements cause significantly lower cybersickness symptoms than continuous locomotion.
Similarly, \citet{adhikari2021hyperjump} and \citet{adhikari_improving_2021} incorporated teleportation into a continuous locomotion technique (controller- or leaning-based). While the user controls simulated self-motion continuously using a given interface, teleportation steps are added periodically in the direction of the continuous locomotion once locomotion speed exceeds a threshold at which cybersickness becomes likely.
The resulting teleportation distance is proportional to joystick deflection or the user's leaning.
The authors reported similar levels of spatial awareness between continuous and noncontinuous movement techniques.

Instead of incorporating teleportation into a continuous locomotion technique, \citet{bhandari_teleportation_2018} took the opposite approach with their \textit{Dash} technique, adding a quick continuous viewport translation to a noncontinuous teleportation.
In a user study, \textit{Dash} was found to significantly reduce the path integration error compared to regular teleportation, while inducing a similar level of cybersickness.
Another approach to creating a continuous avatar representation without causing vection-induced cybersickness is out-of-body locomotion.
This technique allows users to control their avatar from a third-person perspective, and was shown to be effective for both continuous controller-based input~\cite{cmentowski2019outstanding} and noncontinuous teleportation~\cite{griffin2019out}.

Several comparative studies of locomotion techniques regarding self-reported measures have been conducted to determine the most efficient and pleasant form of VR locomotion.
For instance, \citet{Buttussi2019LocomotionInPlace} compared teleportation, joystick-, and leaning-based locomotion techniques in a VR travel task.
They concluded that joysticks and leaning cause similar levels of cybersickness, while teleportation had no significant impact on the users' well-being.
There was no significant difference for the self-reported sense of presence.
Similarly, \citet{xu2017locationMemory} examined distance estimation and spatial awareness in teleportation, joystick-based locomotion and walking-in-place techniques and reported that no significant differences were found.
In contrast, the results of \citet{steinicke2009estimation} indicate that teleportation might induce spatial disorientation and misperception of space and distances.
A comprehensive overview of empirical studies comparing teleportation to both continuous and other noncontinuous locomotion techniques can be found in a review by~\citet{folmer2021teleportation}.\\

As for the related work on avatar visualization, the presented papers on locomotion techniques primarily focus on the experience of a single user, in this case, the moving user.
In a multi-user environment, however, it is equally important to make a user's movements transparent to observers.
Therefore, our goal was to combine the positive effects of continuous locomotion, which can be easily tracked by observers, with the previously discussed benefits of teleportation, which has been shown to reduce cybersickness.
Our resulting technique is called \emph{Stuttered Locomotion} and will be presented in detail in Section~\ref{SubSec:ImplementationStutteredLocomotion}.

\section{Technique Description}
\label{sec:Implementation}
In this section we elaborate on the design and development of the proposed avatar visualization and locomotion techniques' main components.
Our goals were twofold: (i) create an avatar system that provides realistic human-like movement visualizations for noncontinuous locomotion and (ii) provide a cybersickness-reducing mode for continuous locomotion techniques through short-distance teleport steps.
All implementations are available on~\citet{SmartAvatarsGithub}.

\subsection{Smart Avatars - Visualizing Noncontinuous Movements Through Continuous Transitions}
\label{SubSec:ImplementationSmartAvatars}

For noncontinuous locomotion, such as teleportation, conventional avatars would disappear from the initial location and reappear in the target location, which can be a confusing experience for observers in shared VR spaces.
To mask these noncontinuous movements, we developed a novel avatar system that relies on artificial intelligence (AI) agents for continuous human representations.
The core idea is to assign an AI agent to a target user and give them the objective to imitate their assigned user’s tracked real world movement when close-by, and to autonomously navigate to their user when the distance is too long, i.e., when the user teleports.

\subsubsection{Basic Implementation}
To make an avatar’s movement visually as smooth and natural as possible, the locomotion animation has to be perfectly synchronized to the actual translation and rotation of the avatar.
Otherwise, the avatar feet will visually slide over the ground’s surface which risks breaking the illusion of looking at an actual human being.
To achieve this visual synchronization, we utilized Unity’s pathfinding framework `NavMesh' and their animation rigging system `Mecanim'. 
NavMesh defines both walkable surfaces in a given scene and agents which can autonomously navigate and move on these surfaces.
These agents are often used to implement AI-controlled characters and require a single 3D world coordinate as input to perform a pathfinding sequence.
There are also options for off-mesh links to cross gaps between walkable surfaces that require custom behavior, such as jumping between isolated surfaces or climbing ladders.

Smart Avatars can be applied to any animation controller, simple or complex, humanoid or non-humanoid, as long as that controller is compatible with autonomously navigating agents and inverse kinematic (IK) passes.
For our avatar agents we used Unity's standard assets character controller~\cite{unityStandardCharacterController} and a set of root transformation animations obtained from Mixamo~\cite{mixamo}, which will move the avatar entity through 3D space as opposed to merely transforming its assigned mesh.
Rather than having NavMesh move the entity and blending an animation that matches the movement, we take the desired velocity of the NavMesh agent in each frame and interpret this value as input for the Mecanim animator.
Based on this input the animator blends between several animations such as walking forward, leaning and turning.
The result is a root animation that will move the entity in such a way that the visible feet and the entity movement perfectly match.
To summarize, the NavMesh agent determines the required movement, and Mecanim generates a root animation that will satisfy the desired lateral and angular velocity.
This procedure results in smoothly and believably animated agents.
The pathfinding target is constantly updated with the VR camera rig's position, letting the avatar smoothly follow their assigned user.

While the avatar is close to their user, it should imitate their real-world movements and posture.
This is achieved through the combination of a gimbal lock and a second set of animations, primarily consisting of strafe movements.
The gimbal lock constraints yaw rotations of the avatar to the user's current viewing direction, causing the (otherwise unconstrained moving) avatar to lose one degree of freedom.
This constraint is engaged over time to co-incite with the animation blending of the animator blend tree.
When the walking avatar approaches the position of its user from a medium distance and a threshold is passed, it turns quickly, but not abruptly, to face the same direction.
We found that a radius of 2 meters around the camera rig works well as strafing zone.
Inside of this strafing zone, we defined an even smaller zone of 0.5 meters in which we blend in an IK pass to the animator.
Thus, when the avatar and user position overlap substantially, the avatar begins to imitate the user's head and hand movements.

\subsubsection{Transition Types}
Without applying any special effects, the previously described implementation results in a continuous\linebreak \textbf{Walking} animation of the smart avatar whenever its assigned user is teleporting, as illustrated in the supplementary video at 1:05.
The limitation of this technique is that the avatar locomotion only looks realistic up to a certain movement speed.
By limiting the maximum possible movement speed, there can be a considerable discrepancy between avatar and user position when teleporting over long distances.
This could potentially lead to unintentional or inconsistent behavior and interactions between users.
To minimize this positional discrepancy, we developed three long-distance travel visualizations for \emph{Smart Avatars}: \emph{Afterimage}, \emph{Dissolve} and \emph{Foresight}.
When the distance between user and avatar is greater than a predefined threshold, a long-distance travel sequence is initiated to quickly realign avatar and user (see supplementary video at 1:08).

\begin{figure*}[t!]
    \centering
    \includegraphics[width=0.85\textwidth]{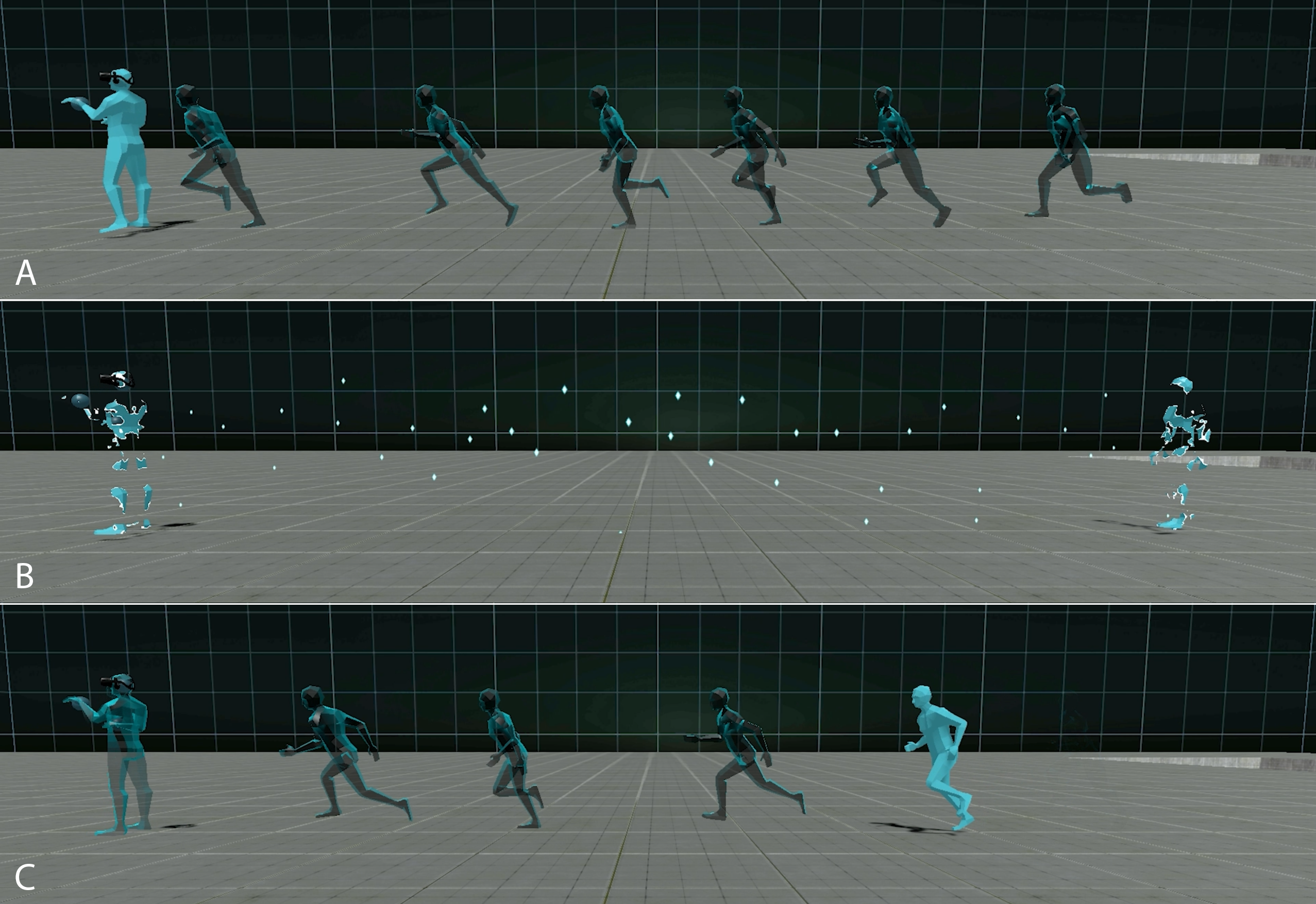}
    \caption[]{The long-distance transition types as seen in the user study's VE: Afterimage (A), Dissolve (B) and Foresight (C).}
    \label{figure:long-distance-travel}
\end{figure*}

\textbf{Afterimage} (cf. Fig.~\ref{figure:long-distance-travel} A) increases the maximum movement speed of the agent tenfold during long-distance travel.
The avatar will additionally leave a trail of ghost-like afterimages that allows observers to visually follow and retrace the path.
In order to create a ghost-like image, we bake the avatar's skinned mesh renderer in its current configuration, apply the resulting mesh to a new entity, and add a transparent texture with fresnel effect that fades out over time.

\textbf{Dissolve} (cf. Fig.~\ref{figure:long-distance-travel} B) likewise creates a copy of the avatar mesh when the sequence is initiated.
Unlike this copy, the original avatar is teleported together with the user and made invisible.
Both original and copy use a dissolve shader with the same noise map; while the original avatar appears over time, the copy disappears over time.
Concurrently a particle system depicts a stream of matter from the initial to the current avatar location.
This gives the impression that the avatar is de- and reconstructed.

\textbf{Foresight} (cf. Fig.~\ref{figure:long-distance-travel} C) uses three agent avatars at the same time for one user:
(i) A ghost-like avatar that is teleported with the user, depicting their current position and imitating their pose without delay.
(ii) An invisible avatar with the same increased movement speed as the \emph{Afterimage} avatar, which leaves a trail of fading ghost-like images.
And (iii) a solid avatar which runs towards the user with default speed.
The trail images are timed in such a way that they disappear once the \emph{Smart Avatar} passes through them.
When triggered, the effect looks like a normal moving user, but with additional ghost-like projections of the avatar's imminent path.

\subsection{Stuttered Locomotion - Mapping Continuous Input To Noncontinuous Movements}
\label{SubSec:ImplementationStutteredLocomotion}

Numerous cybersickness-reducing techniques for VR locomotion are concerned with the user's viewport, for example by limiting their field of view or dynamically blurring their vision~\cite{Fernandes2016FoVReduction,Ang2020GingerVR}.
At the same time, noncontinuous locomotion techniques such as teleportation regularly receive favorable cybersickness ratings when compared to continuous movements like joystick-based locomotion~\cite{freiwald2020VRStrider,Buttussi2019LocomotionInPlace}.
A commonality among these approaches is the limitation of optical flow, which itself was closely related to the level of elicited cybersickness by multiple authors (for an overview, see~\cite{chang2020virtual}).
The approach of reducing optical flow also seems to be applied in several commercial applications, such as the home menus of Steam VR and Oculus, which use discrete rotation steps for joystick-based turning mechanisms~\cite{farmani2020evaluating}. 
From these observations, we conclude that a possible solution to limit the optical flow in continuous locomotion is to decompose a given translation into a series of discrete teleport steps, thus converting a continuous movement into a noncontinuous one.
We call this technique \emph{Stuttered Locomotion}.
\emph{Stuttered Locomotion} can be applied to any continuous motion control method, which we will exemplify with two of these methods, \emph{Joystick} and \emph{PushPull}. In the following, we will first introduce the techniques without the application of \emph{Stuttered Locomotion} before describing the necessary implementational changes for integrating \emph{Stuttered Locomotion}.\\

\textbf{Smooth Joystick} locomotion was included in both the toolkit and the empirical comparison described in Section~\ref{sec:UserStudyInteraction}, as it can be used in regular VR systems without the need for special sensors, and because it is optimal for observers in shared VEs (due to the continuous movements).
In a literature review of 2017, \citet{boletsis2017new} identified controller/joystick as one of the two prevalent continuous locomotion techniques in VR research (besides walking-in-place). The user tilts the joystick on a controller to move in the VE, with the degree of tilt linearly determining the movement speed.\\

\textbf{Stuttered Joystick} locomotion extends the previous motion control by executing a teleport step with a specified step length on the first frame the left joystick is tilted.
While the tilt is continued, another teleport step is triggered each time an internal countdown has passed.
This countdown can be divided through an arbitrary user-selected multiplier or be made dependent on the degree of joystick tilt.
We tested several configurations for step length and fixed/dynamic movement multipliers and derived a number of presets ranging from a focus on efficiency to maximizing comfort.
As part of the provided toolkit, these settings for velocity multipliers, step length, turning steps (i.e., discrete user rotations by a fixed angle), etc. can then be fine-tuned to personal preference by the user in a menu.\\

\textbf{Smooth PushPull} is a hand motion-based locomotion technique with a dynamic velocity multiplier.
It is built upon the Anchor Turning technique described by \citet{freiwald2020VRStrider}, which, from the user’s perspective, lets them drag-and-drop the virtual world around them.
In Anchor Turning, the virtual camera acts as center point of the rotation.
While a drag-and-drop motion is performed, the user is rotated in such a way that the dragging hand stays in a fixed position relative to the virtual world.

For \emph{PushPull}, this technique's principle has been transferred to lateral movements.
While the right hand performs the rotations as described above, the left hand translates the user in a similar manner.
When a drag motion is initiated, the left hand acts as a fix point relative to the world, allowing the user to pull themselves through space.
This metaphor has been described as world-drag before~\cite{VRTK2022Worlddrag}.
The required movement is achieved by saving the initial coordinates of the dragging hand in the user’s local reference frame and comparing this value to the same hand’s current position.
We project this vector onto the plane that is defined by an upwards pointing normal vector to constrain movements to the XZ-plane (in a left-handed coordinate system).
The user’s position is then set to its initial position when the drag motion began, offset by the projected vector.
As both rotation and translation are calculated in the user’s local reference frame, they are independent from each other and can be employed at the same time. 

This approach results in fine movements, which is desirable in terms of achieving a high precision, but unsuitable to cover medium to long distances.
We therefore implemented a dynamic velocity multiplier based on the hand's position on the Y-axis in the user's reference frame.
When the hand is held at chest height or above, the dynamic multiplier is set to $1$, resulting in fine control.
When the hand is held at hip height or lower, a maximum multiplier of $4$, which was determined to result in comfortable and efficient movements in a pilot study, is assumed.
Corresponding multipliers for in-between heights are calculated through linear interpolation.
This dynamic velocity multiplier brings a noteworthy implication:
when the user performs an arm-swing motion starting in front of their body and ending behind their back, their velocity over the movement's duration resembles a parabola with its peak at the halfway point.
Essentially, an arm-swing will result in a short visual dash that eases in and out of the motion, as illustrated in the supplementary video at 2:18.
In summary, \emph{PushPull} is a continuous motion control technique that allows users to be highly flexible with movement options that range from precision to efficiency and dash-like motions.\\

\textbf{Stuttered PushPull} behavior is an extension of the previously described behavior, which requires to add a dead zone around the initial drag position.
Once a certain threshold of aggregated hand movement has been exceeded, a teleportation is initiated with a distance equal to the threshold's value, which we will subsequently refer to as \emph{step length}.
If the vector from initial to current hand position exceeds an exemplary step length of 25cm, a teleport step in the vector's direction is initiated to a point 25cm away from the user's current position.
Then, the process resets and the current hand position is assumed to be the new initial position.
The resulting movement of a user performing an arm-swing motion with \emph{Stuttered PushPull} thus is a series of short-distance teleport steps that cover the same total distance as the same motion with \emph{Smooth PushPull}.
The dynamic velocity multiplier described in the previous section also applies to \emph{Stuttered PushPull}.
However, it had to be implemented in such a way that the step length would remain consistent with the selected user settings.
This can be achieved by dividing the required threshold through the velocity multiplier, as opposed to multiplying the travel distance.

\section{User Study Overview}
\label{sec:UserStudy}
To assess the effectiveness of the introduced techniques, we conducted a multi-stage user study.
The study consisted of an observation section (to compare the avatar visualization techniques) and a first-person interaction task (to compare the locomotion techniques).
The observer and first-person segments will be described separately in the following two sections.

\subsection{Participants and Apparatus}
\label{SubSec:StudyParticipantsAndApparatus}
24 participants (12 male, 12 female) took part in the study.
Age was specified in brackets to comply with general data protection regulation.
12.5\% of participants were in the age group of 18 - 24, 58.3\% in 25 - 34, 16.6\% in 35 - 44, and each 4\% in the groups of 45 - 54, 55 - 64 and 65 - 74.
The mean time per participant was about 20 minutes per study segment, adding up to 40 minutes for the entire two-stage study.
Participants were reminded to take breaks after each condition if they experience cybersickness symptoms.
In accordance with the rules of our Ethics Committee, we assessed ethical concerns using an officially provided basic questionnaire. Based on this assessment (regarding GDPR, experimental protocol, potential harm, cybersickness countermeasures, informed consent, etc.) further consultation with the Ethics Committee was not advised because no potential ethical concerns could be identified for our study.

Unity3D was used to render the scene on an Oculus Quest 2.
The full study was conducted remotely, with detailed instructions for the participants regarding the procedure and requirements for the study environment.


\section{Observer Study}
\label{sec:UserStudyObservation}
The observer part of the user study explores the acceptance and perception of our four \emph{Smart Avatar} transition types, \emph{Walking}, \emph{Afterimage}, \emph{Dissolve}, and \emph{Foresight}.
It simulates a shared VE scenario in which a user (represented by their avatar) performs teleportation while the study participant assumes the role of the observer.
To validate that the advantages of continuous over noncontinuous locomotion demonstrated in related work (e.g., by~\citet{freiwald2021co-presence}) still apply to \emph{Smart Avatars}, we included a \emph{Primitive} condition as a baseline, involving a conventional avatar that disappears and reappears during teleportation.
This conventional avatar consisted of a hovering capsule body with detached hands and head, as this is a common human representation that can be found in commercial products such as Oculus Home.
For the other four levels, we used a low-poly humanoid mesh with natural human locomotion animations, enabled by the \emph{Smart Avatar} technique.
Both types were monocolored.
A joystick-controlled full-body avatar was not included in the comparison as it is, from an observer's perspective, indistinguishable from the \emph{Smart Avatar Walking} transition.
As the study was already fairly long, we did not include additional conditions that could help to further tease apart the relative contributions of avatar fidelity versus transition types in future studies, such as applying \emph{Afterimage}, \emph{Dissolve}, and \emph{Foresight} to a conventional (capsule body) avatar or including a \emph{Smart Avatar} condition without any transitions (thus directly teleporting).

\subsection{Measures}
\label{SubSec:ObservationStudyMeasures}

Of particular interest was the observer's spatial awareness of the avatar user's position and movement intentions.
We expected the default teleportation visualization, where the avatar instantly moves between locations, to be perceived as difficult to anticipate and to visually track.
Since we were not aware of any established questionnaire to validate this hypothesis, we developed a custom questionnaire specifically targeting spatial awareness.
It consisted of three 7-point Likert scales ranging from 1 to 7, asking the participants how strongly they agree with the following statements:
\begin{itemize}
\item ``I had a sense of where the avatar's user currently is.''
\item ``I had a sense of where the avatar's user intends to move.''
\item ``I could reliably point the laser pointer at the avatar.''
\end{itemize}
These values were aggregated in our analysis as the \emph{Spatial Awareness} score.

In addition to spatial awareness, we wanted to investigate whether or not transitions with special effects, such as \emph{Foresight}, have an effect on the perceived pragmatic and hedonic qualities of the avatar.
We expected that the applied visual effects would reduce the humanness of an avatar but make them visually intriguing and easier to predict and track.
To measure the \emph{Pragmatic} and \emph{Hedonic} qualities of the visualizations we employed an English version of the AttrakDiff questionnaire~\cite{hassenzahl2003attrakdiff,AttrakDiff}.
It uses 7-point Likert scales ranging from -3 to +3, asking participants if they would describe the technique rather with the left term (e.g., \emph{unpredictable}, \emph{conventional}) or right term (e.g., \emph{predictable}, \emph{inventive}) of 28 word pairs.
There are four sub-scales, each consisting of 7 word pairs - \emph{Pragmatic Quality} (PQ), \emph{Hedonic Quality - Identity} (HQ-I), \emph{Hedonic Quality - Stimulation} (HQ-S) and \emph{Attractiveness} (ATT).
Here, \emph{Pragmatic Quality} measures usability while \emph{Hedonic Quality} measures emotional reactions, divided by aspects related to the human need for personal development (i.e., stimulation -  e.g., \emph{dull} vs. \emph{engaging}) and for self-expression (i.e., identity - e.g., \emph{isolating} vs. \emph{connecting}).
Additionally, participants were asked to rate their desire to use each visualization technique in a real VR application on a 7-point Likert scale.

\subsection{Hypotheses}

Based on the above-described criteria, the study was designed as within-subject, and the following hypotheses were formed:

\begin{enumerate}[label=\textbf{(H\arabic*)\textsubscript{V}}]
\item Continuous avatar transition techniques are rated higher in spatial awareness than noncontinuous transition techniques.
\item The \emph{Primitive} transition is rated lowest in \emph{Pragmatic} and \emph{Hedonic} qualities.
\item The \emph{Walking} transition appears the most natural and is thus rated highest in \emph{Pragmatic Quality} and \emph{Hedonic Quality - Identity}.
\item The transitions including special effects (i.e., \emph{Afterimage}, \emph{Dissolve}, and \emph{Foresight}) are rated highest in \emph{Hedonic Quality - Stimulation}.
\end{enumerate}

\subsection{Stimuli and Procedure}
\label{SubSec:ObservationStimuliAndProcedue}
In the beginning of a study session, participants filled in a demographic questionnaire and gave their informed consent to participation, followed by a briefing about the nature of the task and the different techniques.

Participants wore an Oculus Quest 2 head-mounted display and were positioned in a mostly empty VE.
In each trial, participants were tasked to observe an avatar moving in a figure eight in front of them.
By having the avatar walk in this pattern, the locomotion can be observed from all sides at different distances.
The figure eight had dimensions of 50 by 10 meters and was oriented orthogonal to the participant, who was standing at a distance of 10 meters from the figure's center point.
The shown avatar animation was a pre-recorded real user sequentially teleporting to 11 points along the figure eight path, over the course of roughly 40 seconds.
Participants held an Oculus controller in their preferred hand that was represented virtually as a laser pointer, which they were asked to point at the avatar at all times.
As we target social situations that naturally involve the user's attention towards another user, for example, a conversation during a meeting or an online shooter game, the laser pointer serves as an abstracted measure to elicit similar attentive behavior as in such practical scenarios.
The laser's color was green while intersecting the avatar, and red when not.
Participants were not embodied beyond visible hands and the attached laser pointer. 

There was one trial for each of the five tested transition types.
The order of trials was randomized by Latin square to counteract learning effects.
After each trial, participants had to rate their experience by filling in the questionnaires described in Section~\ref{SubSec:ObservationStudyMeasures}.
Then, after all trials had concluded, participants were further asked to indicate their preferred avatar visualization.

\newcolumntype{Y}{>{\centering\arraybackslash}X}
\begin{table*}[t!]
    \small
    \centering
    \caption{Means and standard deviations for all measures in the observer study.}
    \label{table_vis}
    \renewcommand{\arraystretch}{1.6}
    \begin{tabularx}{\textwidth}{m{1.8cm}|YY|YY|YY|YY|YY|YY}
        \toprule
        & \multicolumn{2}{c}{\textbf{Pragmatic Quality}} & \multicolumn{2}{c}{\textbf{Hedonic Quality}} & \multicolumn{2}{c}{\textbf{Hedonic Quality}} & \multicolumn{2}{c}{\textbf{Attractiveness}} & \multicolumn{2}{c}{\textbf{Spatial}}
        & \multicolumn{2}{c}{\textbf{Desire}} \\ [-1.7ex]
        & \multicolumn{2}{c}{\textbf{}} & \multicolumn{2}{c}{\textbf{- Identity}} & \multicolumn{2}{c}{\textbf{- Stimulation}} & \multicolumn{2}{c}{\textbf{}} & \multicolumn{2}{c}{\textbf{Awareness}}
        & \multicolumn{2}{c}{\textbf{to use again}}\\
        & \textbf{M} & \textbf{SD} & \textbf{M} & \textbf{SD} & \textbf{M} & \textbf{SD} & \textbf{M} & \textbf{SD} & \textbf{M} & \textbf{SD} & \textbf{M} & \textbf{SD} \\
        \midrule
        \textbf{Primitive} & -0.37 & 1.34 & -0.94 & 1.22 & -1.07 & 1.21 & -0.80 & 1.33 & 3.76 & 1.25 & 2.88 & 1.51 \\
        \textbf{Walking} & 2.07 & 0.96 & 1.30 & 0.88 & -0.09 & 1.18 & 1.40 & 0.90 & 5.79 & 1.00 & 6.17 & 0.94 \\
        \textbf{Afterimage} & 0.31 & 1.25 & 0.96 & 0.94 & 1.42 & 0.86 & 0.82 & 1.24 & 5.18 & 1.02 & 5.00 & 1.58 \\
        \textbf{Dissolve} & 1.02 & 1.15 & 1.36 & 0.80 & 1.19 & 0.79 & 1.45 & 0.97 & 5.89 & 0.68 & 6.00 & 1.08 \\
        \textbf{Foresight} & 0.50 & 1.35 & 1.26 & 0.90 & 1.87 & 0.52 & 1.05 & 1.25 & 5.60 & 1.10 & 5.13 & 1.86 \\
        \bottomrule
    \end{tabularx}
    \renewcommand{\arraystretch}{1.0}
\end{table*}

\subsection{Results}
\label{sec:Results}
\begin{figure*}[t!]
    \centering
    \includegraphics[width=\textwidth]{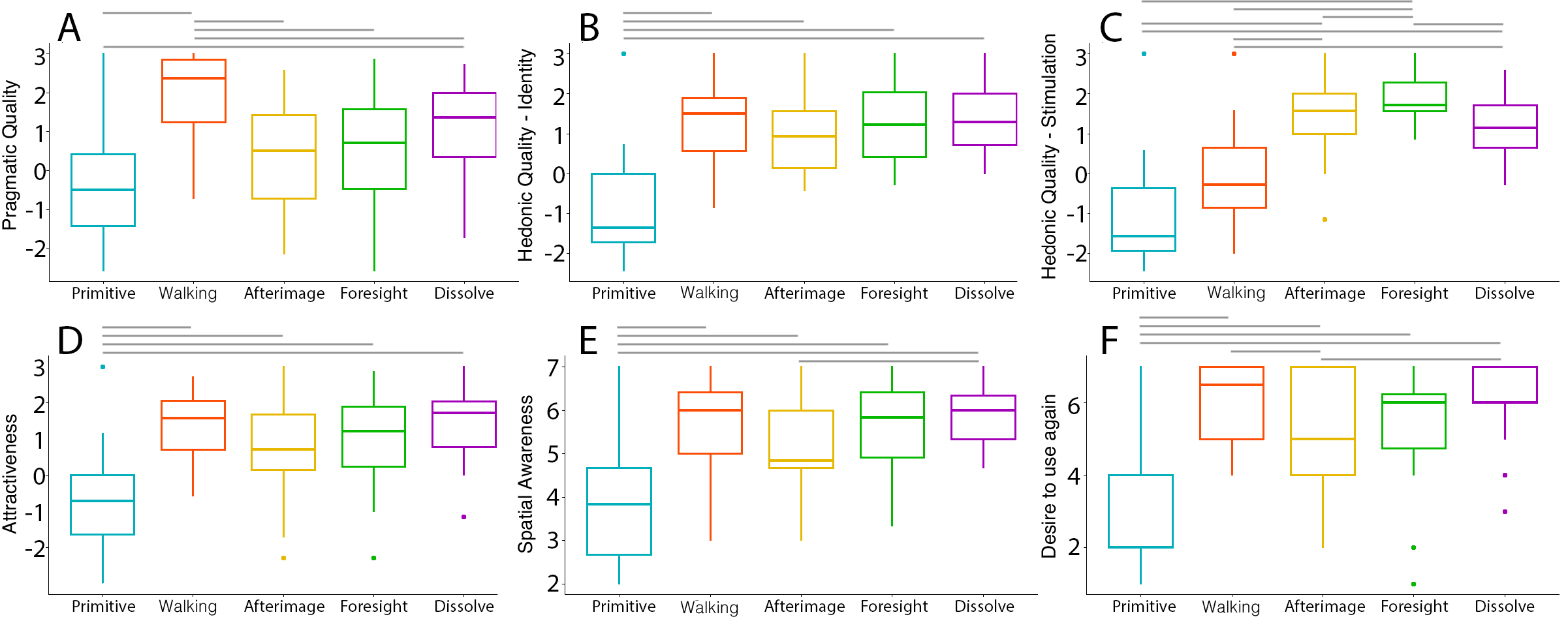}
    \caption[]{Boxplots of the observer study results, indicating means, quartiles, min/max, and outliers. The gray horizontal lines indicate significant differences between conditions in the order of occurrence in Section~\ref{sec:Results}.}
    \label{figure:observationResults}
\end{figure*}

In this section the results of the statistical analysis are presented.
When the Shapiro-Wilk test indicated a normal distribution of residuals, a repeated-measure ANOVA and post-hoc pairwise comparisons with Bonferroni adjustment were used to test for differences between conditions.
Otherwise, the Friedman test and Wilcoxon signed-rank tests with Bonferroni correction were used.
Mauchly’s test did not indicate a violation of the assumption of sphericity for any of the variables.
A 5\% significance level was assumed.
All aggregated questionnaire results are listed in Table~\ref{table_vis} and illustrated in Figure~\ref{figure:observationResults}.
Outliers (shown as dots in Figure~\ref{figure:observationResults}) were included in the statistical analysis because they represent valid opinions for the used subjective measures.

\subsubsection{Pragmatic Quality - cf. Fig.~\ref{figure:observationResults} A}
The ANOVA revealed significant differences between the transition types ($F(4,92)=15.783, p<0.001, \eta_p^2=0.407$).
The \emph{Walking} transition was rated significantly higher than \emph{Primitive} ($p<0.001$), \emph{Afterimage} ($p<0.001$) and \emph{Foresight} ($p=0.001$) as well as \emph{Dissolve} ($p=0.032$).
Additionally, \emph{Dissolve} was rated higher than \emph{Primitive} ($p=0.001$).
The depiction of human gait without artificial effects was therefore rated as the most natural and clearly structured visualization type.
By virtue of not having a fully humanoid shape, it was unsurprising that the primitive avatar received low ratings for humanness.
This, however, is only one item of seven in the  \emph{Pragmatic Quality} category.

\subsubsection{Hedonic Quality - Identity - cf. Fig.~\ref{figure:observationResults} B}
A Friedman test revealed a significant effect of transition type on \emph{Hedonic Quality - Identity} ($\chi^2(4)=41.624, p<0.001$).
A post-hoc analysis with Wilcoxon signed-rank tests was conducted with a Bonferroni correction applied, resulting in a significance level set at $p < 0.01$.
The HQ-I score was significantly lower for the \emph{Primitive} condition in comparison to all other transition types, i.e., \emph{Walking} ($Z=-4.138, p<0.001$), \emph{Afterimage} ($Z=-4.122, p<0.001$), \emph{Dissolve} ($Z=-4.198, p<0.001$) and \emph{Foresight} ($Z=-3.916, p<0.001$).
All variations of the \emph{Smart Avatar} system were rated similarly, highlighting the integrating and connecting properties of naturally and continuously moving full-body avatars.

\subsubsection{Hedonic Quality - Stimulation - cf. Fig.~\ref{figure:observationResults} C}
A Friedman test revealed significant differences between the transition types ($\chi^2(4)=65.502, p<0.001$).
Again, we Bonferroni-adjusted the significance level to $0.01$ and performed multiple Wilcoxon signed-rank tests.
\emph{Foresight} was rated significantly better than all other transitions, i.e.,
\emph{Primitive} ($Z=-4.173, p<0.001$), \emph{Walking} ($Z=-3.958, p<0.001$), \emph{Afterimage} ($Z=-2.994, p=0.003$) and \emph{Dissolve} ($Z=-2.924, p=0.003$).
In addition, \emph{Primitive} was rated significantly lower than \emph{Afterimage} ($Z=-4.198, p<0.001$) and \emph{Dissolve} ($Z=-4.138, p<0.001$).
Also, \emph{Walking} was rated significantly lower than \emph{Afterimage} ($Z=-4.199, p<0.001$) and \emph{Dissolve} ($Z=-3.833, p<0.001$).
The \emph{Foresight} technique was rated highest for the characteristics of \emph{inventive}, \emph{innovative}, \emph{creative} and \emph{captivating}.

\subsubsection{Attractiveness - cf. Fig.~\ref{figure:observationResults} D}
An ANOVA indicated a significant effect of transition type on \emph{Attractiveness}
($F(4,92)=18.742, p<0.001, \eta_p^2=0.449$).
The ATT score was significantly lower for the \emph{Primitive} condition in comparison to all other transition types ($p<0.001$).
All variations of the \emph{Smart Avatar} system were rated similarly, with the same ranking as for the HQ-I score.
Special effects had little influence on the ATT score, signaling that the attractiveness of the locomotion visualization is again directly linked to the continuously moving full-body avatar.

\subsubsection{Spatial Awareness - cf. Fig.~\ref{figure:observationResults} E}
Regarding \emph{Spatial Awareness}, an ANOVA revealed significant differences between transition types
($F(4,92)=20.916, p<0.001, \eta_p^2=0.476$).
The \emph{Spatial Awareness} score was significantly lower for the \emph{Primitive} condition in comparison to all other transition types ($p<0.001$).
Furthermore, \emph{Afterimage} performed significantly worse than \emph{Dissolve} ($p=0.011$).
There were no other significant differences between the \emph{Smart Avatar} variations.
This confirms the findings of~\citet{freiwald2021co-presence}, showing that continuous locomotion is generally easier to interpret and anticipate than noncontinuous locomotion.

\subsubsection{Desire to use again - cf. Fig.~\ref{figure:observationResults} F}
After each trial of the observation experiment, participants were asked to rate their desire to use the depicted visualization in a shared VR space with real users on a scale of 1 to 7.
A Friedman test revealed significant differences between transitions ($\chi^2(4)=44.230, p<0.001$).
According to the post-hoc analysis, the \emph{Primitive} visualization was rated significantly lower than all variations of \emph{Smart Avatars}, i.e., \emph{Walking} ($Z=-4.049, p<0.001$), \emph{Afterimage} ($Z=-3.679, p<0.001$), \emph{Dissolve} ($Z=-4.126, p<0.001$) and \emph{Foresight} ($Z=-3.265, p=0.001$).
Additionally, \emph{Afterimage} got significantly lower scores than \emph{Walking} ($Z=-2.987, p=0.003$) and Dissolve ($Z=-2.857, p=0.004$).
The \emph{Walking} transition received the highest score, however, \emph{Foresight} was rated most often as the favorite avatar type with 37.5\% of votes.
\emph{Dissolve} received 33.33\% of votes, followed by \emph{Walking} with 25\%, \emph{Afterimage} with 4.16\% and \emph{Primitive} with 0\% of votes.

\subsection{Discussion}

In this section we discuss the results of the observer study with regard to the formed hypotheses.

\subsubsection{Higher Spatial Awareness for Continuous Than for Noncontinuous Transitions (H1)\textsubscript{V}}

The \emph{Primitive} transition received the lowest \emph{Spatial Awareness} score, while \emph{Walking} was rated second highest and significantly better than \emph{Primitive}, confirming hypothesis (H1)\textsubscript{V}.
This result suggests that the observer's \emph{Spatial Awareness} score is not positively correlated with the relative time an avatar is displayed at the user's actual position, as this would be highest for \emph{Primitive} avatars.
Therefore, although the displacement between user and \emph{Smart Avatar} can be substantial for long-distance teleports,
the continuous locomotion visualization possibly (falsely) convinced the observers that the user is actually located where the avatar is depicted.
Moreover, the \emph{Dissolve} technique visualizes both start and end point of a long-distance travel simultaneously, while the particle flow implies quick movement, therefore providing users with additional cues to accurately predict the avatar's movement.

\subsubsection{Lowest Pragmatic and Hedonic Quality for Primitive Transition (H2)\textsubscript{V}}

As expected, the \emph{Primitive} transition received the lowest mean ratings for both \emph{Pragmatic} and \emph{Hedonic Quality}, confirming hypothesis (H2)\textsubscript{V}.
Additionally, it received significantly lower \emph{Attractiveness} scores, making it the consistently lowest rated condition in all measures.
It could be argued that the avatar appearance was a major contributing factor to these differences, as it was not identical to the \emph{Smart Avatar} variations.
While similar in color, dimensions and overall complexity, the \emph{Primitive} avatar had no legs.
However, having smoothly animated legs without the need for tracking devices should not be considered an unfair advantage of the \emph{Smart Avatar} conditions, but one of the key technical contributions that set them apart from conventional VR avatars by design.
A comparably animated full-body avatar for the baseline condition would itself have required the application of a technique such as \emph{Smart Avatars}, thus missing the goal of being a baseline to compare the newly created system to.
Moreover, the findings of~\citet{freiwald2021co-presence} suggest that a likely cause of the lower \emph{Attractiveness} scores of the \emph{Primitive} avatar is not its appearance, but the lack of motion animation.
Nonetheless, we are interested in conducting a similar experiment with different avatar visualizations in the future.


Among the \emph{Smart Avatar} transitions, \emph{Afterimage} performed the worst.
From these conditions, it scored lowest in \emph{Spatial Awareness}, \emph{Pragmatic Quality}, \emph{Hedonic Quality - Identity} and \emph{Attractiveness}.
In essence, it is conceptually similar to \emph{Foresight} but inferior in every tested measure.
We therefore cannot recommend using this technique.

\subsubsection{Highest Pragmatic Quality and Hedonic Quality - Identity for Walking Transition (H3)\textsubscript{V}}

As hypothesized, \textit{Walking} received significantly higher \textit{Pragmatic Quality} scores, partially confirming hypothesis (H3)\textsubscript{V}.
Contrary to hypothesis (H3)\textsubscript{V}, \emph{Dissolve} received the top scores for both \emph{Hedonic Quality - Identity} and \emph{Attractiveness}, however without being significantly better than the other \emph{Smart Avatar} transitions.
The graphically elaborate nature of this technique led to a higher score in presentability.
In terms of AttrakDiff word pairs, it was perceived to be \emph{professional}, \emph{stylish} and \emph{premium} in the \emph{Hedonic Quality - Identity} category, and \emph{pleasant}, \emph{appealing} and \emph{motivating} in \emph{Attractiveness}.

\subsubsection{Higher Hedonic Quality - Stimulation for Transitions With Special Effects (H4)\textsubscript{V}}

Our interpretation of the subjective preferences between conditions is that \emph{Foresight} was the most memorable technique, which is supported by receiving the highest scores in the \emph{Hedonic Quality - Stimulation} category of the AttrakDiff questionnaire.
In contrast, the \emph{Walking} transition received the highest mean rating in \emph{Pragmatic Quality} - a measure of humanness and overall usability in terms of being clearly structured and predictable.
In fact, the order of mean scores between \emph{Pragmatic Quality} and desire to reuse a technique were identical.
This strongly suggests that there is a correlation between these two measures.
Overall, \emph{Foresight}, \emph{Afterimage} and \emph{Dissolve} were the strongest transitions for \emph{Hedonic Quality - Stimulation}, confirming hypothesis (H4)\textsubscript{V}.\\


The results conclusively show that the \emph{Smart Avatar} system and the human representations it enables are better suited for shared VR spaces than conventional \emph{Primitive Avatars} from an observer's perspective.
It should be pointed out that it is likely that \emph{Smart Avatars} would be rated similarly to other full-body avatar animation systems during continuous locomotion.
Based on the \emph{Spatial Awareness} results, it stands to reason that observers would not be able to differentiate between those systems and \emph{Smart Avatars}.
The true benefit of using \emph{Smart Avatars} is that the discussed qualities can also be applied to visualizations of noncontinuous locomotion.


\section{First-Person Study}
\label{sec:UserStudyInteraction}
In this section, we describe the locomotion part of the study.
This part was a direct follow-up to the observation study and was thus conducted with the same participants and likewise as within-subject design.
To validate whether \emph{Stuttered Locomotion} indeed reduces the occurrence of cybersickness-related symptoms, we compared it to smooth locomotion for both joystick-based input and our motion-based \emph{PushPull} technique.
Locomotion parameters were chosen to result in comparable maximum movement speeds between \emph{PushPull} and \emph{Joystick} and to reflect average human sprinting.
Based on a pilot study, the maximum \emph{PushPull} velocity multiplier was set to 4, and the maximum Joystick movement speed was 2.5 meters per second (9 km/h).
During the \emph{stuttered} conditions, the step length for both \emph{PushPull} and \emph{Joystick} was set to 0.5 meters and the turning step to 30$^{\circ}$.
\emph{Stuttered Locomotion} works independently of the movement speed as the step length is fixed.

\subsection{Measures}
\label{SubSec:StudyInteractionMeasures}
To inquire the perceived \emph{Efficiency} and \emph{Pleasantness} of the compared techniques, we used one 7-point Likert scale, ranging from 1 to 7, each.

We further employed a modified \emph{Embodiment} questionnaire that covers the dimensions \emph{Body Ownership}, \emph{Agency and Motor Control}, and \emph{Location of the Body} of the Avatar Embodiment Questionnaire of \citet{gonzalez2018avatar}.
The individual questionnaire items are listed in Table~\ref{table_BO}.
Each item was a 7-point Likert scale ranging from 1 to 7, and items 6 and 8 were inverted when calculating the aggregated \emph{Embodiment} score.

\emph{Cybersickness} was rated with a condensed form of the Simulator Sickness Questionnaire (SSQ)~\cite{kennedy1993SSQ}.
Similar to \citet{Fernandes2016FoVReduction}, it consisted of one item for each of the three SSQ sub-scales (nausea, oculomotor and disorientation).
Every item was represented by a 5-point Likert scale ranging from 0 (not affected) to 4 (strongly affected).

\emph{Presence} was rated using the Slater-Usoh-Steed presence questionnaire (SUSP)~\cite{slater1994depth} consisting of 6 items, mainly concerned with the participant's memory of the virtual space being similar to their perception of the real world.
Each item was a 7-point Likert scale ranging from 0 to 6.

\emph{Usability} was rated using the System Usability Scale (SUS)~\cite{brooke1996sus}.
It covers aspects such as system complexity, learnability and ease of use.
Using ten 5-point Likert scale items each ranging from 1 to 5, a total SUS score between 0 and 100 is calculated.

\subsection{Hypotheses}

Based on the above-described criteria the following hypotheses were formed:

\begin{enumerate}[label=\textbf{(H\arabic*)\textsubscript{L}}]
\item \emph{PushPull}'s hand-based movement is directly linked to the user's proprioception and thus receives lower \emph{Cybersickness} scores than \emph{Joystick}.
\item \emph{PushPull} is physically more demanding than \emph{Joystick}, leading to lower \emph{Usability}.
\item \emph{Stuttered Locomotion} has less optical flow and will receive lower \emph{Cybersickness} scores than \emph{Smooth Locomotion}.
\item \emph{Stuttered Locomotion} is unnatural and will be perceived as less \emph{Pleasant} than \emph{Smooth Locomotion}.
\end{enumerate}

\subsection{Stimuli and Procedure}
\label{SubSec:StimuliAndProcedue}
This experiment directly followed the observation task.
At this point participants were reminded they could take breaks between sub-sections.
An initial locomotion tutorial prepared the participants to use the \emph{Joystick} and \emph{PushPull} methods by having them reach certain target areas in their vicinity.

After the tutorial concluded, participants were asked to indicate their current level of cybersickness by filling in the condensed SSQ.
During the experiment, the participants' task was to find and slice oversized fruits with a sword, and afterwards return to their starting position.
There were three sets of three fruits each, arranged in such a way that the participant had to move forward to reach the first set, turn approximately 90$^{\circ}$ left to find the second set, turn approximately 180$^{\circ}$ to see the third set, and finally turn approximately 90$^{\circ}$ right to return to their highlighted starting point.
The next set of fruits was only activated when the previous one was successfully cut.
The distance between the starting point and each set of fruits was 5 meters.
A black vignette from \emph{VR Tunneling Pro}~\cite{VRTunnelingPro} with default values was faded in during movement or rotation regardless of locomotion technique, which is in line with current state-of-the-art VR experiences~\cite{AlZayer2019FoVRestriction}.
In each trial only one combination of input method (\emph{Joystick} / \emph{PushPull}) and stuttering (\emph{Smooth} / \emph{Stuttered}) was available, while the order of trials was randomized by Latin square.
After each trial, participants were asked to fill in the questionnaires described in Section~\ref{SubSec:StudyInteractionMeasures}.

\newcolumntype{Y}{>{\centering\arraybackslash}X}
\begin{table*}[t!]
    \small
    \centering
    \caption{Means and standard deviations for all measures in the first-person study.}
    \label{table_loc}
    \renewcommand{\arraystretch}{1.6}
    \begin{tabularx}{\textwidth}{m{2.4cm}|YY|YY|YY|YY|YY|YY}
        \toprule
        & \multicolumn{2}{c}{\textbf{Efficiency}} & \multicolumn{2}{c}{\textbf{Pleasantness}} & \multicolumn{2}{c}{\textbf{Embodiment}} & \multicolumn{2}{c}{\textbf{Cybersickness}} & \multicolumn{2}{c}{\textbf{Presence}} &
        \multicolumn{2}{c}{\textbf{Usability}} \\
        & \textbf{M} & \textbf{SD} & \textbf{M} & \textbf{SD} & \textbf{M} & \textbf{SD} & \textbf{M} & \textbf{SD} & \textbf{M} & \textbf{SD} & \textbf{M} & \textbf{SD} \\
        \midrule
        \textbf{Smooth Joystick} & 4.71 & 1.90 & 3.92 & 2.15 & 5.45 & 1.13 & 10.29 & 8.91 & 2.92 & 1.11 & 75.52 & 17.55 \\
        \textbf{Stuttered Joystick} & 4.13 & 1.57 & 4.21 & 1.56 & 5.22 & 1.05 & -3.12 & 8.30 & 3.17 & 1.08 & 74.79 & 17.37 \\
        \textbf{Smooth PushPull} & 4.21 & 1.67 & 4.29 & 1.37 & 5.51 & 1.07 & 0.16 & 5.89 & 3.22 & 1.26 & 66.04 & 16.23 \\
        \textbf{Stuttered PushPull} & 3.46 & 1.59 & 4.38 & 1.97 & 5.20 & 1.27 & -3.90 & 8.21 & 3.25 & 1.13 & 56.35 & 24.25 \\
        \bottomrule
    \end{tabularx}
    \renewcommand{\arraystretch}{1.0}
\end{table*}

\subsection{Results}
\label{sec:StudyInteractionResults}
\begin{figure*}[t!]
    \centering
    \includegraphics[width=\textwidth]{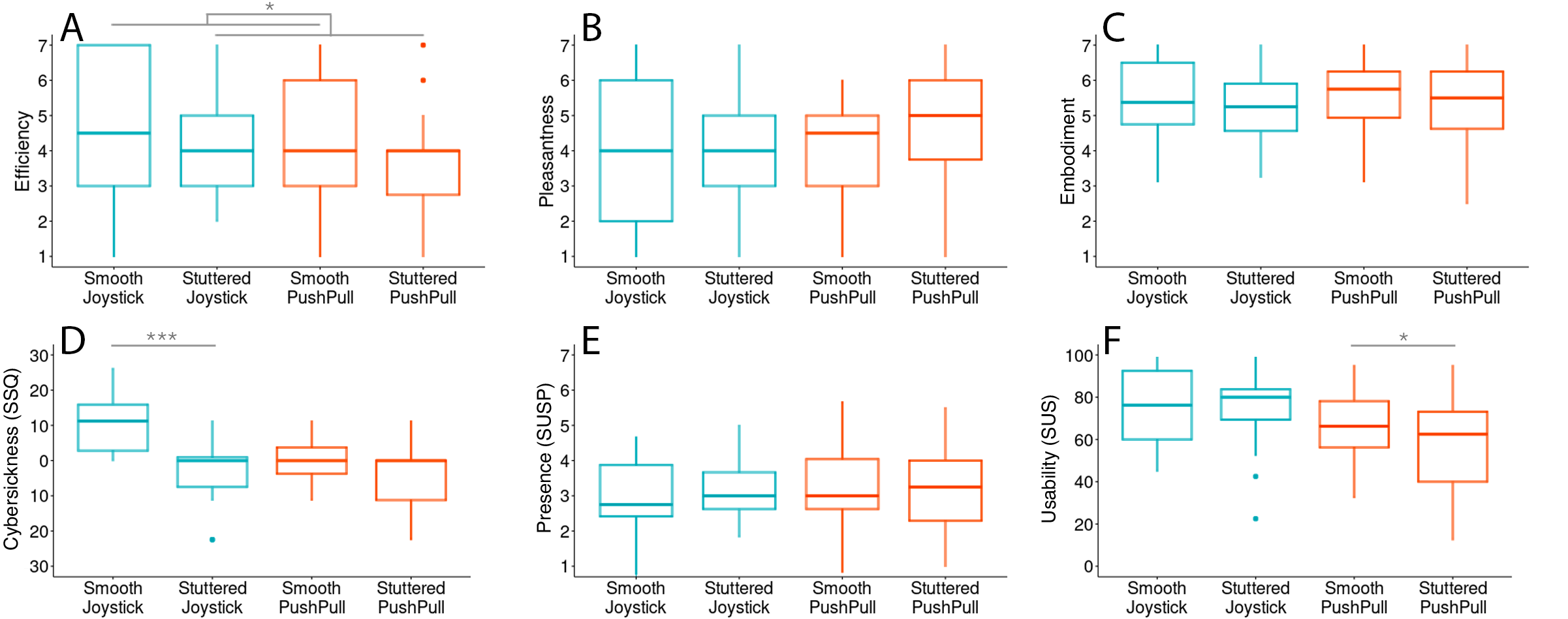}
    \caption[]{Boxplots of the first-person study results, indicating means, quartiles, min/max, and outliers. The gray horizontal lines indicate significant differences between conditions (for significant interactions, only the significant simple main effects are illustrated).}
    \label{figure:interactionResults}
\end{figure*}

We considered the two factors \emph{Input Method} (\emph{Joystick} / \emph{PushPull}) and \emph{Stuttering}~(\emph{Smooth} / \emph{Stuttered}), and therefore performed a two-way ANOVA with a 5\% significance level for each of the dependent variables.
For \emph{Pleasantness} and \emph{Embodiment}, an inspection of both histograms and Q-Q plots showed a small negative skewness ($-0.280$ and $-0.510$, respectively) and kurtosis ($-0.701$ and $-0.448$, respectively) of the residuals, however, the ANOVA has been shown to be robust against such mild deviations from the normal distribution~\cite{norman2010likert}.
All questionnaire results are listed in Table~\ref{table_loc} and illustrated in Figure~\ref{figure:interactionResults}.
As argued for the observer study, outliers were included in the statistical analysis.

\subsubsection{Efficiency - cf. Fig.~\ref{figure:interactionResults} A}
A two-way ANOVA revealed a significant main effect of \emph{Stuttering} on self-reported \emph{Efficiency} ($F(1,23)=7.472, p=0.012, \eta_p^2=0.245$).
\emph{Stuttered} locomotion ($M=3.79, SD=1.60$) was perceived as significantly less efficient than \emph{Smooth} locomotion ($M=4.46, SD=1.79$).
We did not find a significant difference between \emph{Joystick} and \emph{PushPull} locomotion.

\subsubsection{Pleasantness - cf. Fig.~\ref{figure:interactionResults} B}
No significant main or interaction effects of \emph{Input Method} and \emph{Stuttering} on \emph{Pleasantness} were found.

\subsubsection{Embodiment - cf. Fig.~\ref{figure:interactionResults} C}
As for \emph{Pleasantness}, we also did not find any significant effects of the two factors on \emph{Embodiment}.

\renewcommand{\arraystretch}{1.3}
\begin{table}[h!]
\centering
\resizebox{\columnwidth}{!}{%
\begin{tabular}{|c|l|}
\hline
Code & Item \\
\hline
E1 & I caused the movement and cutting actions. \\
E2 & I performed the movement and cutting actions myself. \\
E3 & The visible avatar represented me. \\
E4 & I felt like I could control the avatar as if it was my own body. \\
E5 & I felt as if the avatar was my body. \\
E6 & I felt out of my body. \\
E7 & I felt as if my body was located where I saw the avatar. \\
E8 & I felt like the avatar was someone else. \\
\hline
\end{tabular}}
\caption{The \emph{Embodiment} questionnaire based on \citet{gonzalez2018avatar}. Participants were asked how strongly they agree with these statements.}
\label{table_BO}
\end{table}
\renewcommand{\arraystretch}{1.0}

\subsubsection{Cybersickness - cf. Fig.~\ref{figure:interactionResults} D}
For the SSQ, we compared the differences from the questionnaire results gathered before and after every trial (POST-PRE).
Therefore, positive results indicate an increase in \emph{Cybersickness}, while negative results indicate a decrease or return to normal upon usage of the method.
A two-way ANOVA revealed a significant spreading interaction between \emph{Input Method} and \emph{Stuttering} ($F(1,23)=10.147, p=0.004, \eta_p^2=0.306$).
We also found significant main effects of \emph{Input Method} ($F(1,23)=17.149, p<0.001, \eta_p^2=0.427$) and \emph{Stuttering} ($F(1,23)=12.963, p=0.002, \eta_p^2=0.360$) on \emph{Cybersickness}, however, they only have limited conclusiveness in the presence of a significant interaction.
We therefore performed a follow-up simple main effects analysis with Sidak-adjusted comparisons, which indicated that for \emph{Joystick} input, \emph{Cybersickness} was rated $13.402$ points higher for \emph{Smooth} locomotion than for \emph{Stuttered} locomotion ($p < 0.001$).
For \emph{PushPull} input, no significant difference between \emph{Smooth} and \emph{Stuttered} was found in the simple main effects analysis ($p=0.108$).

\subsubsection{Presence - cf. Fig.~\ref{figure:interactionResults} E}
For the self-reported sense of \emph{Presence}, all conditions received similar ratings, and therefore no significant differences were found.

\subsubsection{Usability - cf. Fig.~\ref{figure:interactionResults} F}
Regarding self-reported \emph{Usability}, a two-way ANOVA revealed a significant main effect of \emph{Input Method} ($F(1,23)=8.659, p=0.007, \eta_p^2=0.274$) as well as a significant interaction effect between \emph{Input Method} and \emph{Stuttering} ($F(1,23)=6.386, p=0.019, \eta_p^2=0.217$).
We conducted a simple main effects analysis with Sidak adjustment and found a significant difference between \emph{Smooth PushPull} and \emph{Stuttered PushPull} ($p=0.011$), with the latter scoring $9.687$ points worse on average.
For \emph{Joystick} input, no significant simple main effect was found ($p=0.864$).

\subsection{Discussion}
\label{sec:DiscussionLoc}
In this section we discuss the results of the first-person study with regard to the formed hypotheses.

\subsubsection{Lower Cybersickness for PushPull Than for Joystick (H1)\textsubscript{L}}

The analysis of the SSQ revealed that \emph{PushPull} indeed caused fewer occurrences of cybersickness symptoms than \emph{Joystick}, thus confirming hypothesis (H1)\textsubscript{L}.
The major benefit of motion-based locomotion, the reduction of cybersickness, also applies to \emph{PushPull}, making it an attractive choice for regular usage.
This applies in particular to \emph{PushPull} in combination with \emph{Stuttered Locomotion}, as discussed in Section~\ref{sec:discussionH3}.
For Joystick-based locomotion, the comparatively higher mean score for cybersickness was primarily due to the \emph{Smooth} condition, while \emph{Stuttered Joystick} received similar scores to both \emph{PushPull} conditions.

\subsubsection{Lower Usability for PushPull Than for Joystick (H2)\textsubscript{L}}

\emph{PushPull}'s \emph{Usability} score was significantly lower than that of \emph{Joystick}, confirming hypothesis (H2)\textsubscript{L}.
Participants reported that \emph{PushPull} required more time to learn and master.
As it is similar to the drag-and-drop metaphor, we expected it to be natural and easy to learn, which did not appear to be the case.
A potential factor for this might be the fact that motion-based input was a novel concept to many participants.
This had no effect on the perceived \emph{Efficiency} of the technique, however.

At the same time, \emph{PushPull} and \emph{Joystick} received similar ratings for \emph{Pleasantness}, \emph{Embodiment} and sense of \emph{Presence}.
Regarding those measures, other works found significant differences between teleportation and joysticks as well as motion-based locomotion and leaning~\cite{freiwald2020VRStrider,Buttussi2019LocomotionInPlace}, but no such effects could be found between \emph{PushPull} and \emph{Joystick} in this study.


\subsubsection{Lower Cybersickness for Stuttered Locomotion Than for Smooth Locomotion (H3)\textsubscript{L}}
\label{sec:discussionH3}

Regarding \emph{Stuttered Locomotion}, we predicted a reduction in \emph{Cybersickness} due to fewer instances of optical flow, a hypothesis that is supported by the results of our analysis, in particular for \emph{Joystick} locomotion.
Thus, hypothesis (H3)\textsubscript{L} was confirmed.
These results are similar to those of \citet{farmani2020evaluating}, who implemented discrete forward/backward teleport steps of each 1 meter that can be triggered via mouse buttons.
However, \emph{Stuttered Locomotion} expands on this idea by being compatible to virtually any continuous locomotion technique, while supporting movements in any direction with a multitude of available user settings.
In the context of teleport steps, a participant came forward with an interesting analogy.
They compared their perception of \emph{Stuttered Locomotion} to the unique gait of birds like pigeons and chickens.
During ground movement, those animals keep their heads in a stable position while the body moves and abruptly thrust their head forward to the next stable position.
The most likely theory for why these birds bob their heads is for the same reason that we move our eyes around – to stabilize the image of their surroundings while in motion~\cite{frost1978optokinetic}.
This is indeed similar to the \emph{Stuttered Locomotion} approach to reducing \emph{Cybersickness} by limiting optical flow, motivating further investigations in the future.

\subsubsection{Lower Pleasantness for Stuttered Locomotion Than for Smooth Locomotion (H4)\textsubscript{L}}

We found no significant differences for the measure of \emph{Pleasantness} between \emph{Stuttered} and \emph{Smooth Locomotion}.
Hence, \emph{Stuttered Locomotion} was received more positively than anticipated and hypothesis (H4)\textsubscript{L} was rejected.
Although the analysis suggests that stuttering has no disadvantages over smooth locomotion in this regard, individual participants reported that they perceived this technique to be unpleasant.
When analyzing the interaction effect for \emph{Usability}, we indeed noticed that the combination of \emph{Stuttered Locomotion} and \emph{PushPull} was rated strikingly unfavorable.
We suspect that this is due to the initial threshold that has to be passed in order for the first translation to take place.
Until the threshold is passed, there is no visual feedback on whether or not the technique works correctly, whereas there is immediate feedback for both joystick configurations.
One solution to this problem is to visualize the vector from initial to current hand position in form of an arrow inside a sphere with the radius of the required threshold.
When the arrow touches the wall of the sphere, a teleport step is initiated and the arrow resets.
If this extension is considered, our recommendation based on the presented user study is to offer \emph{Stuttered Locomotion} as an option whenever continuous locomotion is employed.

\section{Limitations and Future Work}

For the presented multi-stage study, we intended to create a minimal scenario for an initial evaluation of the proposed techniques focused on the subjective user experience, while ensuring maximum control over potential confounding factors.
The obtained positive empirical results warrant follow-up studies to evaluate the generalizability to (i) more natural virtual environments, (ii) more than two users, and (iii) specific interaction tasks.

Regarding the latter, we only measured whether observers had the impression that they knew where the other user was, which can be the case even if there is a considerable difference between the user's actual position and the visualized position of their smart avatar.
Due to this natural mismatch, an analysis of the accuracy of the laser pointer (introduced to ensure the user's directed attention) would not be meaningful, as it does not imply improved visualization effectiveness.
Instead, other objective measures could be included in future studies to examine the extent to which the subjectively assumed position and the actual position match, and how this affects interaction quality in shared environments.
For example, the use of collaborative assembly tasks (e.g., as in ~\citet{pan2017impact}) would allow the introduction of additional objective measures such as task completion time or error rate.


Further potential for future investigation lies in the transitions themselves.
Each explored transition focuses on visualizing different aspects of the user's locomotion, thus covering a wide range of possibilities.
While teleportation only indicates the target position and walking visualizes a user's movement, Dissolve simultaneously visualizes the start and target position plus the movement. AfterImage and Foresight show the path rather than the movement itself, either with a focus on the target or the starting point. We hypothesize that pragmatic quality and spatial awareness mainly depend on these visualized aspects of locomotion rather than the actual appearance of the transition. Our initial results with exemplary transitions support the usefulness of the underlying Smart Avatars concept, therefore warranting future research of other visualizations.

A concrete extension of the transition selection would be to include a condition with a fully tracked primitive avatar.
For the presented study, we used current state-of-the-art VR headsets (Oculus Quest 2) that do not yet incorporate full-body tracking. 
Therefore, such setups prohibit the use of leg animation.
In a future controlled lab study, an external tracking system could be used for validating the current results with a fully tracked primitive avatar.

Finally, the current study did not assess how participants would perceive \emph{Smart Avatar} transitions from their own perspective.
Instead of seeing their own avatar only at the actual destination, users could be shown the same transitions as observers when they look behind them or via mirrors in the scene.
Evaluating such a representation of one's avatar during movements in terms of measures such as embodiment and perceived presence would be an interesting extension of the work presented, constituting a goal for future work.

\section{Conclusion}
\label{sec:Conclusion} 

In this paper, we presented two approaches for improving shared virtual experiences by reducing confusion of observers while a user is (potentially noncontinuously) moving through the VE.
We addressed this challenge (i) from the observer perspective, by visualizing noncontinuous movements of users via continuous avatar transitions, and (ii) from a first-person perspective, by introducing a locomotion mode that allows mapping continuous input to short-interval movements with reduced cybersickness.

In terms of transitions, we introduced and compared four \emph{Smart Avatar} techniques, which use different continuous visualizations for noncontinuous movements.
All the techniques have in common that the avatar imitates the assigned user's (head and arm) movements when nearby, but switches to an autonomous navigation as soon as the user performs a noncontinuous movement in the VE (e.g., teleportation).
In a user study, observers rated \emph{Smart Avatars} significantly higher in self-reported \emph{Spatial Awareness}, \emph{Pragmatic} and \emph{Hedonic} qualities as well as \emph{Attractiveness} when compared to conventional avatars.

A second approach to ensure that observers can follow a user's avatar movements would be to utilize continuous locomotion techniques, such as joystick-based input.
However, since this has been shown to induce higher levels of cybersickness compared to teleportation, we proposed a concept called \emph{Stuttered Locomotion}.
It involves the decomposition of a continuous movement into short teleport steps that are easier to track by observers than a single teleport over a long distance, but at the same time can lead to reduced cybersickness levels of the moving user.
These hypothesized positive effects were confirmed in a second user study in which \emph{Stuttered Locomotion} was applied to two continuous locomotion techniques.
This benefit was more substantial for the joystick-based input, possibly due to \emph{PushPull}'s demonstrated property of causing less cybersickness than joystick movement overall.
On the downside, it was also found that \emph{Stuttered Locomotion} was perceived as significantly less efficient than smooth locomotion.

The techniques of \emph{Smart Avatars} and \emph{Stuttered Locomotion} for both joysticks and \emph{PushPull} were bundled into a toolkit that was made openly available for other researchers to use and extend.\\

While both approaches can separately contribute to a smoother experience for multiple moving users in a shared VE, they can also be combined.
Since teleportation has been previously shown to increase spatial disorientation not only for the observer but also for the moving user~\cite{steinicke2009estimation}, \emph{Stuttered Locomotion} could provide an alternative, especially for traveling short distances. 
In fact, the intermittent short-range teleports that were added on top of continuous locomotion in \emph{HyperJump} by Adhikari et al. \cite{adhikari_improving_2021,adhikari2021hyperjump,riecke_hyperjumping_2022} did not impair spatial orientation, which suggests that \emph{Stuttered Locomotion} might similarly be able to reduce cybersickness without significantly impairing users' spatial orientation. 
Compared to related techniques like \emph{Dashing} \cite{bhandari_teleportation_2018} or \emph{Viewpoint Snapping} \cite{farmani2020evaluating,farmani_viewpoint_2018}, that can also provide repeated jumps, \emph{Stuttered Locomotion} has the advantage of not needing to be manually triggered, thus presumably reducing cognitive load. 
By adding the \emph{Smart Avatar} system on top, stuttered motion can be smoothed out to simulate a fully continuous avatar movement for the observers.
Different nuances of interplay between moving users and observers are the subject of future studies.

\section*{Acknowledgments}

This work was supported by the German Federal Ministry of Education and Research (BMBF).

\balance{}

\bibliographystyle{ACM-Reference-Format}
\bibliography{bibliography}


\begin{thebibliography}{54}


\ifx \showCODEN    \undefined \def \showCODEN     #1{\unskip}     \fi
\ifx \showDOI      \undefined \def \showDOI       #1{#1}\fi
\ifx \showISBNx    \undefined \def \showISBNx     #1{\unskip}     \fi
\ifx \showISBNxiii \undefined \def \showISBNxiii  #1{\unskip}     \fi
\ifx \showISSN     \undefined \def \showISSN      #1{\unskip}     \fi
\ifx \showLCCN     \undefined \def \showLCCN      #1{\unskip}     \fi
\ifx \shownote     \undefined \def \shownote      #1{#1}          \fi
\ifx \showarticletitle \undefined \def \showarticletitle #1{#1}   \fi
\ifx \showURL      \undefined \def \showURL       {\relax}        \fi
\providecommand\bibfield[2]{#2}
\providecommand\bibinfo[2]{#2}
\providecommand\natexlab[1]{#1}
\providecommand\showeprint[2][]{arXiv:#2}

\bibitem[Adcock et~al\mbox{.}(2013)]%
        {adcock2013remotefusion}
\bibfield{author}{\bibinfo{person}{Matt Adcock}, \bibinfo{person}{Stuart
  Anderson}, {and} \bibinfo{person}{Bruce Thomas}.}
  \bibinfo{year}{2013}\natexlab{}.
\newblock \showarticletitle{RemoteFusion: real time depth camera fusion for
  remote collaboration on physical tasks}. In
  \bibinfo{booktitle}{\emph{Proceedings of the 12th ACM SIGGRAPH international
  conference on virtual-reality continuum and its applications in industry}}.
  \bibinfo{pages}{235--242}.
\newblock


\bibitem[Adhikari(2021)]%
        {adhikari_improving_2021}
\bibfield{author}{\bibinfo{person}{Ashu Adhikari}.}
  \bibinfo{year}{2021}\natexlab{}.
\newblock \emph{\bibinfo{title}{Improving {Spatial} {Orientation} in {Virtual}
  {Reality} with {Leaning}-based {Interfaces}}}.
\newblock \bibinfo{thesistype}{Master's\ thesis}. \bibinfo{school}{Simon Fraser
  University}, \bibinfo{address}{Surrey, BC, Canada}.
\newblock
\urldef\tempurl%
\url{https://summit.sfu.ca/item/21858}
\showURL{%
\tempurl}


\bibitem[Adhikari et~al\mbox{.}(2021)]%
        {adhikari2021hyperjump}
\bibfield{author}{\bibinfo{person}{Ashu Adhikari}, \bibinfo{person}{Daniel
  Zielasko}, \bibinfo{person}{Alexander Bretin}, \bibinfo{person}{Markus
  von~der Heyde}, \bibinfo{person}{Ernst Kruijff}, {and}
  \bibinfo{person}{Bernhard~E Riecke}.} \bibinfo{year}{2021}\natexlab{}.
\newblock \showarticletitle{Integrating Continuous and Teleporting VR
  Locomotion into a Seamless" HyperJump" Paradigm}. In
  \bibinfo{booktitle}{\emph{2021 IEEE Conference on Virtual Reality and 3D User
  Interfaces Abstracts and Workshops (VRW)}}. IEEE, \bibinfo{pages}{370--372}.
\newblock


\bibitem[{Adobe Systems Incorporated}(2022)]%
        {mixamo}
\bibfield{author}{\bibinfo{person}{{Adobe Systems Incorporated}}.}
  \bibinfo{year}{2022}\natexlab{}.
\newblock \bibinfo{booktitle}{\emph{Mixamo}}.
\newblock
\urldef\tempurl%
\url{https://www.mixamo.com/}
\showURL{%
\tempurl}


\bibitem[Al~Zayer et~al\mbox{.}(2019)]%
        {AlZayer2019FoVRestriction}
\bibfield{author}{\bibinfo{person}{Majed Al~Zayer}, \bibinfo{person}{Isayas~B.
  Adhanom}, \bibinfo{person}{Paul MacNeilage}, {and} \bibinfo{person}{Eelke
  Folmer}.} \bibinfo{year}{2019}\natexlab{}.
\newblock \showarticletitle{The Effect of Field-of-View Restriction on Sex Bias
  in VR Sickness and Spatial Navigation Performance}. In
  \bibinfo{booktitle}{\emph{Proceedings of the 2019 CHI Conference on Human
  Factors in Computing Systems}} (Glasgow, Scotland Uk)
  \emph{(\bibinfo{series}{CHI '19})}. \bibinfo{publisher}{ACM},
  \bibinfo{address}{New York, NY, USA}, Article \bibinfo{articleno}{354},
  \bibinfo{numpages}{12}~pages.
\newblock
\showISBNx{978-1-4503-5970-2}


\bibitem[Ang and Quarles(2020)]%
        {Ang2020GingerVR}
\bibfield{author}{\bibinfo{person}{Samuel Ang} {and} \bibinfo{person}{John
  Quarles}.} \bibinfo{year}{2020}\natexlab{}.
\newblock \showarticletitle{GingerVR: An Open Source Repository of
  Cybersickness Reduction Techniques for Unity}. In
  \bibinfo{booktitle}{\emph{2020 IEEE Conference on Virtual Reality and 3D User
  Interfaces Abstracts and Workshops (VRW)}}. \bibinfo{pages}{460--463}.
\newblock
\urldef\tempurl%
\url{https://doi.org/10.1109/VRW50115.2020.00097}
\showDOI{\tempurl}


\bibitem[Bhandari et~al\mbox{.}(2018)]%
        {bhandari_teleportation_2018}
\bibfield{author}{\bibinfo{person}{Jiwan Bhandari}, \bibinfo{person}{Paul
  MacNeilage}, {and} \bibinfo{person}{Eelke Folmer}.}
  \bibinfo{year}{2018}\natexlab{}.
\newblock \showarticletitle{Teleportation without {Spatial} {Disorientation}
  using {Optical} {Flow} {Cues}}. In \bibinfo{booktitle}{\emph{Graphics
  {Interface} 2018}}. \bibinfo{address}{Toronto, Canada}, \bibinfo{pages}{153
  -- 158}.
\newblock


\bibitem[Boletsis(2017)]%
        {boletsis2017new}
\bibfield{author}{\bibinfo{person}{Costas Boletsis}.}
  \bibinfo{year}{2017}\natexlab{}.
\newblock \showarticletitle{The new era of virtual reality locomotion: A
  systematic literature review of techniques and a proposed typology}.
\newblock \bibinfo{journal}{\emph{Multimodal Technologies and Interaction}}
  \bibinfo{volume}{1}, \bibinfo{number}{4} (\bibinfo{year}{2017}),
  \bibinfo{pages}{24}.
\newblock


\bibitem[Boletsis and Cedergren(2019)]%
        {boletsis2019vr}
\bibfield{author}{\bibinfo{person}{Costas Boletsis} {and}
  \bibinfo{person}{Jarl~Erik Cedergren}.} \bibinfo{year}{2019}\natexlab{}.
\newblock \showarticletitle{Vr locomotion in the new era of virtual reality: An
  empirical comparison of prevalent techniques}.
\newblock \bibinfo{journal}{\emph{Advances in Human-Computer Interaction}}
  \bibinfo{volume}{2019} (\bibinfo{year}{2019}).
\newblock


\bibitem[Brooke et~al\mbox{.}(1996)]%
        {brooke1996sus}
\bibfield{author}{\bibinfo{person}{John Brooke} {et~al\mbox{.}}}
  \bibinfo{year}{1996}\natexlab{}.
\newblock \showarticletitle{SUS-A quick and dirty usability scale}.
\newblock \bibinfo{journal}{\emph{Usability evaluation in industry}}
  \bibinfo{volume}{189}, \bibinfo{number}{194} (\bibinfo{year}{1996}),
  \bibinfo{pages}{4--7}.
\newblock


\bibitem[Buttussi and Chittaro(2019)]%
        {Buttussi2019LocomotionInPlace}
\bibfield{author}{\bibinfo{person}{Fabio Buttussi} {and} \bibinfo{person}{Luca
  Chittaro}.} \bibinfo{year}{2019}\natexlab{}.
\newblock \showarticletitle{Locomotion in Place in Virtual Reality: A
  Comparative Evaluation of Joystick, Teleport, and Leaning}.
\newblock \bibinfo{journal}{\emph{IEEE Transactions on Visualization and
  Computer Graphics}} (\bibinfo{year}{2019}).
\newblock


\bibitem[Cardoso(2016)]%
        {cardoso2016comparison}
\bibfield{author}{\bibinfo{person}{Jorge~CS Cardoso}.}
  \bibinfo{year}{2016}\natexlab{}.
\newblock \showarticletitle{Comparison of gesture, gamepad, and gaze-based
  locomotion for VR worlds}. In \bibinfo{booktitle}{\emph{Proceedings of the
  22nd ACM conference on virtual reality software and technology}}.
  \bibinfo{pages}{319--320}.
\newblock


\bibitem[Casaneuva(2001)]%
        {casaneuva2001presence}
\bibfield{author}{\bibinfo{person}{Juan~S Casaneuva}.}
  \bibinfo{year}{2001}\natexlab{}.
\newblock \emph{\bibinfo{title}{Presence and co-presence in collaborative
  virtual environments}}.
\newblock \bibinfo{thesistype}{Ph.\,D. Dissertation}.
  \bibinfo{school}{University of Cape Town}.
\newblock


\bibitem[Chang et~al\mbox{.}(2020)]%
        {chang2020virtual}
\bibfield{author}{\bibinfo{person}{Eunhee Chang}, \bibinfo{person}{Hyun~Taek
  Kim}, {and} \bibinfo{person}{Byounghyun Yoo}.}
  \bibinfo{year}{2020}\natexlab{}.
\newblock \showarticletitle{Virtual reality sickness: a review of causes and
  measurements}.
\newblock \bibinfo{journal}{\emph{International Journal of Human--Computer
  Interaction}} \bibinfo{volume}{36}, \bibinfo{number}{17}
  (\bibinfo{year}{2020}), \bibinfo{pages}{1658--1682}.
\newblock


\bibitem[Cmentowski et~al\mbox{.}(2019)]%
        {cmentowski2019outstanding}
\bibfield{author}{\bibinfo{person}{Sebastian Cmentowski},
  \bibinfo{person}{Andrey Krekhov}, {and} \bibinfo{person}{Jens Kr{\"u}ger}.}
  \bibinfo{year}{2019}\natexlab{}.
\newblock \showarticletitle{Outstanding: A multi-perspective travel approach
  for virtual reality games}. In \bibinfo{booktitle}{\emph{Proceedings of the
  annual symposium on computer-human interaction in play}}.
  \bibinfo{pages}{287--299}.
\newblock


\bibitem[Farmani and Teather(2018)]%
        {farmani_viewpoint_2018}
\bibfield{author}{\bibinfo{person}{Yasin Farmani} {and}
  \bibinfo{person}{Robert~J. Teather}.} \bibinfo{year}{2018}\natexlab{}.
\newblock \showarticletitle{Viewpoint {Snapping} to {Reduce} {Cybersickness} in
  {Virtual} {Reality}}. In \bibinfo{booktitle}{\emph{Proceedings of the 44th
  {Graphics} {Interface} {Conference}}} \emph{(\bibinfo{series}{{GI} '18})}.
  \bibinfo{publisher}{Canadian Human-Computer Communications Society},
  \bibinfo{address}{Toronto, Canada}, \bibinfo{pages}{168--175}.
\newblock
\showISBNx{978-0-9947868-3-8}
\urldef\tempurl%
\url{https://doi.org/10.20380/GI2018.23}
\showDOI{\tempurl}


\bibitem[Farmani and Teather(2020)]%
        {farmani2020evaluating}
\bibfield{author}{\bibinfo{person}{Yasin Farmani} {and}
  \bibinfo{person}{Robert~J Teather}.} \bibinfo{year}{2020}\natexlab{}.
\newblock \showarticletitle{Evaluating discrete viewpoint control to reduce
  cybersickness in virtual reality}.
\newblock \bibinfo{journal}{\emph{Virtual Reality}} \bibinfo{volume}{24},
  \bibinfo{number}{4} (\bibinfo{year}{2020}), \bibinfo{pages}{645--664}.
\newblock


\bibitem[Fernandes and Feiner(2016)]%
        {Fernandes2016FoVReduction}
\bibfield{author}{\bibinfo{person}{Ajoy~S Fernandes} {and}
  \bibinfo{person}{Steven~K. Feiner}.} \bibinfo{year}{2016}\natexlab{}.
\newblock \showarticletitle{Combating VR sickness through subtle dynamic
  field-of-view modification}. In \bibinfo{booktitle}{\emph{2016 IEEE Symposium
  on 3D User Interfaces (3DUI)}}. \bibinfo{pages}{201--210}.
\newblock
\urldef\tempurl%
\url{https://doi.org/10.1109/3DUI.2016.7460053}
\showDOI{\tempurl}


\bibitem[Ferracani et~al\mbox{.}(2016)]%
        {ferracani2016locomotion}
\bibfield{author}{\bibinfo{person}{Andrea Ferracani}, \bibinfo{person}{Daniele
  Pezzatini}, \bibinfo{person}{Jacopo Bianchini}, \bibinfo{person}{Gianmarco
  Biscini}, {and} \bibinfo{person}{Alberto Del~Bimbo}.}
  \bibinfo{year}{2016}\natexlab{}.
\newblock \showarticletitle{Locomotion by natural gestures for immersive
  virtual environments}. In \bibinfo{booktitle}{\emph{Proceedings of the 1st
  international workshop on multimedia alternate realities}}.
  \bibinfo{pages}{21--24}.
\newblock


\bibitem[Folmer et~al\mbox{.}(2021)]%
        {folmer2021teleportation}
\bibfield{author}{\bibinfo{person}{Eelke Folmer}, \bibinfo{person}{Isayas~Berhe
  Adhanom}, {and} \bibinfo{person}{Aniruddha Prithul}.}
  \bibinfo{year}{2021}\natexlab{}.
\newblock \showarticletitle{Teleportation in Virtual Reality; A Mini-Review}.
\newblock \bibinfo{journal}{\emph{Frontiers in Virtual Reality}}
  (\bibinfo{year}{2021}), \bibinfo{pages}{138}.
\newblock


\bibitem[Freiwald et~al\mbox{.}(2020)]%
        {freiwald2020VRStrider}
\bibfield{author}{\bibinfo{person}{Jann~Philipp Freiwald},
  \bibinfo{person}{Oscar Ariza}, \bibinfo{person}{Omar Janeh}, {and}
  \bibinfo{person}{Frank Steinicke}.} \bibinfo{year}{2020}\natexlab{}.
\newblock \showarticletitle{Walking by Cycling: A Novel In-Place Locomotion
  User Interface for Seated Virtual Reality Experiences}. In
  \bibinfo{booktitle}{\emph{Proceedings of the 2020 CHI Conference on Human
  Factors in Computing Systems}}. \bibinfo{pages}{1--12}.
\newblock


\bibitem[Freiwald et~al\mbox{.}(2021)]%
        {freiwald2021co-presence}
\bibfield{author}{\bibinfo{person}{Jann~Philipp Freiwald},
  \bibinfo{person}{Julius Schenke}, \bibinfo{person}{Nale Lehmann-Willenbrock},
  {and} \bibinfo{person}{Frank Steinicke}.} \bibinfo{year}{2021}\natexlab{}.
\newblock \showarticletitle{Effects of Avatar Appearance and Locomotion on
  Co-Presence in Virtual Reality Collaborations}. In
  \bibinfo{booktitle}{\emph{Mensch Und Computer 2021}} (Ingolstadt, Germany)
  \emph{(\bibinfo{series}{MuC '21})}. \bibinfo{publisher}{Association for
  Computing Machinery}, \bibinfo{address}{New York, NY, USA},
  \bibinfo{pages}{393–401}.
\newblock
\showISBNx{9781450386456}
\urldef\tempurl%
\url{https://doi.org/10.1145/3473856.3473870}
\showDOI{\tempurl}


\bibitem[Frost(1978)]%
        {frost1978optokinetic}
\bibfield{author}{\bibinfo{person}{Barrie Frost}.}
  \bibinfo{year}{1978}\natexlab{}.
\newblock \showarticletitle{The optokinetic basis of head-bobbing in the
  pigeon}.
\newblock \bibinfo{journal}{\emph{Journal of Experimental Biology}}
  \bibinfo{volume}{74}, \bibinfo{number}{1} (\bibinfo{year}{1978}),
  \bibinfo{pages}{187--195}.
\newblock


\bibitem[{GitHub}(2022)]%
        {SmartAvatarsGithub}
\bibfield{author}{\bibinfo{person}{{GitHub}}.} \bibinfo{year}{2022}\natexlab{}.
\newblock \bibinfo{booktitle}{\emph{Smart Avatars GitHub Repository}}.
\newblock
\urldef\tempurl%
\url{https://github.com/uhhhci/TheNonContinuousToolkit}
\showURL{%
\tempurl}


\bibitem[Gonzalez-Franco and Peck(2018)]%
        {gonzalez2018avatar}
\bibfield{author}{\bibinfo{person}{Mar Gonzalez-Franco} {and}
  \bibinfo{person}{Tabitha~C Peck}.} \bibinfo{year}{2018}\natexlab{}.
\newblock \showarticletitle{Avatar embodiment. towards a standardized
  questionnaire}.
\newblock \bibinfo{journal}{\emph{Frontiers in Robotics and AI}}
  \bibinfo{volume}{5} (\bibinfo{year}{2018}), \bibinfo{pages}{74}.
\newblock


\bibitem[Griffin and Folmer(2019)]%
        {griffin2019out}
\bibfield{author}{\bibinfo{person}{Nathan~Navarro Griffin} {and}
  \bibinfo{person}{Eelke Folmer}.} \bibinfo{year}{2019}\natexlab{}.
\newblock \showarticletitle{Out-of-body locomotion: Vectionless navigation with
  a continuous avatar representation}. In \bibinfo{booktitle}{\emph{25th ACM
  Symposium on Virtual Reality Software and Technology}}.
  \bibinfo{pages}{1--8}.
\newblock


\bibitem[Grochow et~al\mbox{.}(2004)]%
        {grochow2004style}
\bibfield{author}{\bibinfo{person}{Keith Grochow}, \bibinfo{person}{Steven~L
  Martin}, \bibinfo{person}{Aaron Hertzmann}, {and} \bibinfo{person}{Zoran
  Popovi{\'c}}.} \bibinfo{year}{2004}\natexlab{}.
\newblock \showarticletitle{Style-based inverse kinematics}.
\newblock In \bibinfo{booktitle}{\emph{ACM SIGGRAPH 2004 Papers}}.
  \bibinfo{pages}{522--531}.
\newblock


\bibitem[Harris et~al\mbox{.}(2014)]%
        {harris2014human}
\bibfield{author}{\bibinfo{person}{Alyssa Harris}, \bibinfo{person}{Kevin
  Nguyen}, \bibinfo{person}{Preston~Tunnell Wilson}, \bibinfo{person}{Matthew
  Jackoski}, {and} \bibinfo{person}{Betsy Williams}.}
  \bibinfo{year}{2014}\natexlab{}.
\newblock \showarticletitle{Human joystick: Wii-leaning to translate in large
  virtual environments}. In \bibinfo{booktitle}{\emph{Proceedings of the 13th
  ACM SIGGRAPH International Conference on Virtual-Reality Continuum and its
  Applications in Industry}}. \bibinfo{pages}{231--234}.
\newblock


\bibitem[Hassenzahl et~al\mbox{.}(2003)]%
        {hassenzahl2003attrakdiff}
\bibfield{author}{\bibinfo{person}{Marc Hassenzahl}, \bibinfo{person}{Michael
  Burmester}, {and} \bibinfo{person}{Franz Koller}.}
  \bibinfo{year}{2003}\natexlab{}.
\newblock \showarticletitle{AttrakDiff: Ein Fragebogen zur Messung
  wahrgenommener hedonischer und pragmatischer Qualit{\"a}t [AttrakDiff: A
  questionnaire to measure perceived hedonic and pragmatic quality]}.
\newblock In \bibinfo{booktitle}{\emph{Mensch \& Computer 2003}}.
  \bibinfo{publisher}{Springer}, \bibinfo{pages}{187--196}.
\newblock


\bibitem[Kennedy et~al\mbox{.}(1993)]%
        {kennedy1993SSQ}
\bibfield{author}{\bibinfo{person}{Robert~S. Kennedy},
  \bibinfo{person}{Norman~E. Lane}, \bibinfo{person}{Kevin~S. Berbaum}, {and}
  \bibinfo{person}{Michael~G. Lilienthal}.} \bibinfo{year}{1993}\natexlab{}.
\newblock \showarticletitle{Simulator sickness questionnaire: An enhanced
  method for quantifying simulator sickness}.
\newblock \bibinfo{journal}{\emph{The international journal of aviation
  psychology}} \bibinfo{volume}{3}, \bibinfo{number}{3} (\bibinfo{year}{1993}),
  \bibinfo{pages}{203--220}.
\newblock


\bibitem[LaViola~Jr(2000)]%
        {laviola2000discussion}
\bibfield{author}{\bibinfo{person}{Joseph~J LaViola~Jr}.}
  \bibinfo{year}{2000}\natexlab{}.
\newblock \showarticletitle{A discussion of cybersickness in virtual
  environments}.
\newblock \bibinfo{journal}{\emph{ACM Sigchi Bulletin}} \bibinfo{volume}{32},
  \bibinfo{number}{1} (\bibinfo{year}{2000}), \bibinfo{pages}{47--56}.
\newblock


\bibitem[McCullough et~al\mbox{.}(2015)]%
        {mccullough2015myo}
\bibfield{author}{\bibinfo{person}{Morgan McCullough}, \bibinfo{person}{Hong
  Xu}, \bibinfo{person}{Joel Michelson}, \bibinfo{person}{Matthew Jackoski},
  \bibinfo{person}{Wyatt Pease}, \bibinfo{person}{William Cobb},
  \bibinfo{person}{William Kalescky}, \bibinfo{person}{Joshua Ladd}, {and}
  \bibinfo{person}{Betsy Williams}.} \bibinfo{year}{2015}\natexlab{}.
\newblock \showarticletitle{Myo arm: swinging to explore a VE}. In
  \bibinfo{booktitle}{\emph{Proceedings of the ACM SIGGRAPH Symposium on
  Applied Perception}}. \bibinfo{pages}{107--113}.
\newblock


\bibitem[Mohler et~al\mbox{.}(2010)]%
        {Mohler2010AvatarDistanceEstimation}
\bibfield{author}{\bibinfo{person}{Betty~J. Mohler}, \bibinfo{person}{Sarah~H.
  Creem-Regehr}, \bibinfo{person}{William~B. Thompson}, {and}
  \bibinfo{person}{Heinrich~H. Buelthoff}.} \bibinfo{year}{2010}\natexlab{}.
\newblock \showarticletitle{The Effect of Viewing a Self-Avatar on Distance
  Judgments in an HMD-Based Virtual Environment}.
\newblock \bibinfo{journal}{\emph{Presence: Teleoperators and Virtual
  Environments}} \bibinfo{volume}{19}, \bibinfo{number}{3}
  (\bibinfo{year}{2010}), \bibinfo{pages}{230--242}.
\newblock


\bibitem[Norman(2010)]%
        {norman2010likert}
\bibfield{author}{\bibinfo{person}{Geoff Norman}.}
  \bibinfo{year}{2010}\natexlab{}.
\newblock \showarticletitle{Likert scales, levels of measurement and the
  “laws” of statistics}.
\newblock \bibinfo{journal}{\emph{Advances in health sciences education}}
  \bibinfo{volume}{15}, \bibinfo{number}{5} (\bibinfo{year}{2010}),
  \bibinfo{pages}{625--632}.
\newblock


\bibitem[Orts-Escolano et~al\mbox{.}(2016)]%
        {orts2016holoportation}
\bibfield{author}{\bibinfo{person}{Sergio Orts-Escolano},
  \bibinfo{person}{Christoph Rhemann}, \bibinfo{person}{Sean Fanello},
  \bibinfo{person}{Wayne Chang}, \bibinfo{person}{Adarsh Kowdle},
  \bibinfo{person}{Yury Degtyarev}, \bibinfo{person}{David Kim},
  \bibinfo{person}{Philip~L Davidson}, \bibinfo{person}{Sameh Khamis},
  \bibinfo{person}{Mingsong Dou}, {et~al\mbox{.}}}
  \bibinfo{year}{2016}\natexlab{}.
\newblock \showarticletitle{Holoportation: Virtual 3d teleportation in
  real-time}. In \bibinfo{booktitle}{\emph{Proceedings of the 29th annual
  symposium on user interface software and technology}}.
  \bibinfo{pages}{741--754}.
\newblock


\bibitem[Pan and Steed(2017)]%
        {pan2017impact}
\bibfield{author}{\bibinfo{person}{Ye Pan} {and} \bibinfo{person}{Anthony
  Steed}.} \bibinfo{year}{2017}\natexlab{}.
\newblock \showarticletitle{The impact of self-avatars on trust and
  collaboration in shared virtual environments}.
\newblock \bibinfo{journal}{\emph{PloS one}} \bibinfo{volume}{12},
  \bibinfo{number}{12} (\bibinfo{year}{2017}), \bibinfo{pages}{e0189078}.
\newblock


\bibitem[Pan and Steed(2019)]%
        {pan2019foot}
\bibfield{author}{\bibinfo{person}{Ye Pan} {and} \bibinfo{person}{Anthony
  Steed}.} \bibinfo{year}{2019}\natexlab{}.
\newblock \showarticletitle{How foot tracking matters: The impact of an
  animated self-avatar on interaction, embodiment and presence in shared
  virtual environments}.
\newblock \bibinfo{journal}{\emph{Frontiers in Robotics and AI}}
  \bibinfo{volume}{6} (\bibinfo{year}{2019}), \bibinfo{pages}{104}.
\newblock


\bibitem[Perez-Marcos et~al\mbox{.}(2012)]%
        {perez2012bodyconnected}
\bibfield{author}{\bibinfo{person}{Daniel Perez-Marcos},
  \bibinfo{person}{Maria~V Sanchez-Vives}, {and} \bibinfo{person}{Mel Slater}.}
  \bibinfo{year}{2012}\natexlab{}.
\newblock \showarticletitle{Is my hand connected to my body? The impact of body
  continuity and arm alignment on the virtual hand illusion}.
\newblock \bibinfo{journal}{\emph{Cognitive neurodynamics}}
  \bibinfo{volume}{6}, \bibinfo{number}{4} (\bibinfo{year}{2012}),
  \bibinfo{pages}{295--305}.
\newblock


\bibitem[Piumsomboon et~al\mbox{.}(2018)]%
        {Piumsomboon2018mini-me}
\bibfield{author}{\bibinfo{person}{Thammathip Piumsomboon},
  \bibinfo{person}{Gun~A. Lee}, \bibinfo{person}{Jonathon~D. Hart},
  \bibinfo{person}{Barrett Ens}, \bibinfo{person}{Robert~W. Lindeman},
  \bibinfo{person}{Bruce~H. Thomas}, {and} \bibinfo{person}{Mark
  Billinghurst}.} \bibinfo{year}{2018}\natexlab{}.
\newblock \bibinfo{booktitle}{\emph{Mini-Me: An Adaptive Avatar for Mixed
  Reality Remote Collaboration}}.
\newblock \bibinfo{publisher}{Association for Computing Machinery},
  \bibinfo{address}{New York, NY, USA}, \bibinfo{pages}{1–13}.
\newblock
\showISBNx{9781450356206}
\urldef\tempurl%
\url{https://doi.org/10.1145/3173574.3173620}
\showURL{%
\tempurl}


\bibitem[Riecke et~al\mbox{.}(2022)]%
        {riecke_hyperjumping_2022}
\bibfield{author}{\bibinfo{person}{Bernhard~E Riecke}, \bibinfo{person}{David
  Clement}, \bibinfo{person}{Denise Quesnel}, \bibinfo{person}{Ashu Adhikari},
  \bibinfo{person}{Daniel Zielasko}, {and} \bibinfo{person}{Markus von~der
  Heyde}.} \bibinfo{year}{2022}\natexlab{}.
\newblock \showarticletitle{{HyperJumping} in {Virtual} {Vancouver}:
  {Combating} {Motion} {Sickness} by {Merging} {Teleporting} and {Continuous}
  {VR} {Locomotion} in an {Embodied} {Hands}-{Free} {VR} {Flying} {Paradigm}}.
  In \bibinfo{booktitle}{\emph{Siggraph '22 {Immersive} {Pavilion}}}.
  \bibinfo{publisher}{ACM}, \bibinfo{address}{Vancouver, BC, Canada},
  \bibinfo{pages}{1--2}.
\newblock
\showISBNx{978-1-4503-9369-0}
\urldef\tempurl%
\url{https://doi.org/10.1145/3532834.3536211}
\showDOI{\tempurl}


\bibitem[{Sigtrap Ltd.}(2022)]%
        {VRTunnelingPro}
\bibfield{author}{\bibinfo{person}{{Sigtrap Ltd.}}}
  \bibinfo{year}{2022}\natexlab{}.
\newblock \bibinfo{booktitle}{\emph{VR Tunneling Pro}}.
\newblock
\urldef\tempurl%
\url{http://www.sigtrapgames.com/VrTunnellingPro/html}
\showURL{%
\tempurl}


\bibitem[Slater et~al\mbox{.}(1994)]%
        {slater1994depth}
\bibfield{author}{\bibinfo{person}{Mel Slater}, \bibinfo{person}{Martin Usoh},
  {and} \bibinfo{person}{Anthony Steed}.} \bibinfo{year}{1994}\natexlab{}.
\newblock \showarticletitle{Depth of presence in virtual environments}.
\newblock \bibinfo{journal}{\emph{Presence: Teleoperators \& Virtual
  Environments}} \bibinfo{volume}{3}, \bibinfo{number}{2}
  (\bibinfo{year}{1994}), \bibinfo{pages}{130--144}.
\newblock


\bibitem[Sra and Schmandt(2015)]%
        {sra2015metaspace}
\bibfield{author}{\bibinfo{person}{Misha Sra} {and} \bibinfo{person}{Chris
  Schmandt}.} \bibinfo{year}{2015}\natexlab{}.
\newblock \showarticletitle{Metaspace: Full-body tracking for immersive
  multiperson virtual reality}. In \bibinfo{booktitle}{\emph{Adjunct
  Proceedings of the 28th Annual ACM Symposium on User Interface Software \&
  Technology}}. \bibinfo{pages}{47--48}.
\newblock


\bibitem[Steed et~al\mbox{.}(2016)]%
        {steed2016wild}
\bibfield{author}{\bibinfo{person}{Anthony Steed}, \bibinfo{person}{Sebastian
  Frlston}, \bibinfo{person}{Maria~Murcia Lopez}, \bibinfo{person}{Jason
  Drummond}, \bibinfo{person}{Ye Pan}, {and} \bibinfo{person}{David Swapp}.}
  \bibinfo{year}{2016}\natexlab{}.
\newblock \showarticletitle{An ‘in the wild’experiment on presence and
  embodiment using consumer virtual reality equipment}.
\newblock \bibinfo{journal}{\emph{IEEE transactions on visualization and
  computer graphics}} \bibinfo{volume}{22}, \bibinfo{number}{4}
  (\bibinfo{year}{2016}), \bibinfo{pages}{1406--1414}.
\newblock


\bibitem[Steinicke et~al\mbox{.}(2009)]%
        {steinicke2009estimation}
\bibfield{author}{\bibinfo{person}{Frank Steinicke}, \bibinfo{person}{Gerd
  Bruder}, \bibinfo{person}{Jason Jerald}, \bibinfo{person}{Harald Frenz},
  {and} \bibinfo{person}{Markus Lappe}.} \bibinfo{year}{2009}\natexlab{}.
\newblock \showarticletitle{Estimation of detection thresholds for redirected
  walking techniques}.
\newblock \bibinfo{journal}{\emph{IEEE transactions on visualization and
  computer graphics}} \bibinfo{volume}{16}, \bibinfo{number}{1}
  (\bibinfo{year}{2009}), \bibinfo{pages}{17--27}.
\newblock


\bibitem[Teo et~al\mbox{.}(2019)]%
        {teo2019mixed}
\bibfield{author}{\bibinfo{person}{Theophilus Teo}, \bibinfo{person}{Louise
  Lawrence}, \bibinfo{person}{Gun~A Lee}, \bibinfo{person}{Mark Billinghurst},
  {and} \bibinfo{person}{Matt Adcock}.} \bibinfo{year}{2019}\natexlab{}.
\newblock \showarticletitle{Mixed reality remote collaboration combining 360
  video and 3d reconstruction}. In \bibinfo{booktitle}{\emph{Proceedings of the
  2019 CHI conference on human factors in computing systems}}.
  \bibinfo{pages}{1--14}.
\newblock


\bibitem[Tregillus and Folmer(2016)]%
        {tregillus2016vr}
\bibfield{author}{\bibinfo{person}{Sam Tregillus} {and} \bibinfo{person}{Eelke
  Folmer}.} \bibinfo{year}{2016}\natexlab{}.
\newblock \showarticletitle{Vr-step: Walking-in-place using inertial sensing
  for hands free navigation in mobile vr environments}. In
  \bibinfo{booktitle}{\emph{Proceedings of the 2016 CHI Conference on Human
  Factors in Computing Systems}}. \bibinfo{pages}{1250--1255}.
\newblock


\bibitem[{Unity Technologies}(2019)]%
        {unityStandardCharacterController}
\bibfield{author}{\bibinfo{person}{{Unity Technologies}}.}
  \bibinfo{year}{2019}\natexlab{}.
\newblock \bibinfo{booktitle}{\emph{Unity Standard Assets: Character Package}}.
\newblock
\urldef\tempurl%
\url{https://github.com/Unity-Technologies/Standard-Assets-Characters}
\showURL{%
\tempurl}


\bibitem[{User Interface Design}(2022)]%
        {AttrakDiff}
\bibfield{author}{\bibinfo{person}{{User Interface Design}}.}
  \bibinfo{year}{2022}\natexlab{}.
\newblock \bibinfo{booktitle}{\emph{AttrakDiff}}.
\newblock
\urldef\tempurl%
\url{http://www.attrakdiff.de/index-en.html/}
\showURL{%
\tempurl}


\bibitem[{VRChat}(2022)]%
        {VRChat2022Avatars30}
\bibfield{author}{\bibinfo{person}{{VRChat}}.} \bibinfo{year}{2022}\natexlab{}.
\newblock \bibinfo{booktitle}{\emph{Avatars 3.0}}.
\newblock
\urldef\tempurl%
\url{https://docs.vrchat.com/docs/avatars-30}
\showURL{%
\tempurl}


\bibitem[{VRTK}(2022)]%
        {VRTK2022Worlddrag}
\bibfield{author}{\bibinfo{person}{{VRTK}}.} \bibinfo{year}{2022}\natexlab{}.
\newblock \bibinfo{booktitle}{\emph{VRTK DragWorld}}.
\newblock
\urldef\tempurl%
\url{https://vrtoolkit.readme.io/docs/vrtk_dragworld}
\showURL{%
\tempurl}


\bibitem[Waltemate et~al\mbox{.}(2018)]%
        {Waltemate2018PersonalizedAvatars}
\bibfield{author}{\bibinfo{person}{Thomas Waltemate}, \bibinfo{person}{Dominik
  Gall}, \bibinfo{person}{Daniel Roth}, \bibinfo{person}{Mario Botsch}, {and}
  \bibinfo{person}{Marc~Erich Latoschik}.} \bibinfo{year}{2018}\natexlab{}.
\newblock \showarticletitle{The Impact of Avatar Personalization and Immersion
  on Virtual Body Ownership, Presence, and Emotional Response}.
\newblock \bibinfo{journal}{\emph{IEEE Transactions on Visualization and
  Computer Graphics}} \bibinfo{volume}{24}, \bibinfo{number}{4}
  (\bibinfo{date}{apr} \bibinfo{year}{2018}), \bibinfo{pages}{1643–1652}.
\newblock
\showISSN{1077-2626}
\urldef\tempurl%
\url{https://doi.org/10.1109/TVCG.2018.2794629}
\showDOI{\tempurl}


\bibitem[Wilson et~al\mbox{.}(2016)]%
        {wilson2016vr}
\bibfield{author}{\bibinfo{person}{Preston~Tunnell Wilson},
  \bibinfo{person}{William Kalescky}, \bibinfo{person}{Ansel MacLaughlin},
  {and} \bibinfo{person}{Betsy Williams}.} \bibinfo{year}{2016}\natexlab{}.
\newblock \showarticletitle{VR locomotion: walking> walking in place> arm
  swinging}. In \bibinfo{booktitle}{\emph{Proceedings of the 15th ACM SIGGRAPH
  Conference on Virtual-Reality Continuum and Its Applications in
  Industry-Volume 1}}. \bibinfo{pages}{243--249}.
\newblock


\bibitem[Xu et~al\mbox{.}(2017)]%
        {xu2017locationMemory}
\bibfield{author}{\bibinfo{person}{Mengxin Xu}, \bibinfo{person}{Mar{\'\i}a
  Murcia-L{\'o}pez}, {and} \bibinfo{person}{Anthony Steed}.}
  \bibinfo{year}{2017}\natexlab{}.
\newblock \showarticletitle{Object location memory error in virtual and real
  environments}. In \bibinfo{booktitle}{\emph{2017 IEEE Virtual Reality (VR)}}.
  IEEE, \bibinfo{pages}{315--316}.
\newblock


\end{thebibliography}





\end{document}